\documentclass{article}

\usepackage{a4wide}
\usepackage{url}
\usepackage{graphicx}
\usepackage{float}
\usepackage{varioref}
\usepackage{amssymb,amsmath}
\usepackage[round]{natbib}

\begin{document}

\title{Supervised, semi-supervised and unsupervised inference of gene regulatory networks }   
       
\author{Stefan R. Maetschke\,$^{1}$, Piyush B. Madhamshettiwar\,$^{1}$, 
     \\Melissa J. Davis\,$^{1}$ and Mark A. Ragan\,$^{1,2}$}        
\date{}
\maketitle

$^{1}$The University of Queensland, Institute for Molecular Bioscience 

$^{2}$Australian Research Council Centre of Excellence in Bioinformatics

\vspace{1cm}

\begin{abstract}

Inference of gene regulatory network from expression data is a challenging task. Many methods have been developed to this purpose but a comprehensive evaluation that covers unsupervised,
semi-supervised and supervised methods, and provides guidelines for their practical application, is lacking.

We performed an extensive evaluation of inference methods on simulated expression data. The results reveal very low prediction accuracies for unsupervised techniques with the notable exception of the z-score method on knock-out data. In all other cases the supervised approach achieved the highest accuracies and even in a semi-supervised setting with small numbers of only positive samples, outperformed the unsupervised techniques.

\end{abstract}

\newpage
\section{Introduction}

Mapping the topology of gene regulatory networks is a central problem in systems biology. The regulatory architecture controlling gene expression also controls consequent cellular behavior such as development, differentiation, homeostasis and response to stimuli, while deregulation of these networks has been implicated in oncogenesis and tumor progression \citep{Peer2011}.  Experimental methods based e.g. on chromatin immunoprecepitation, DNaseI hypersensitivity or protein-binding assays are capable of determining the nature of gene regulation in a given system, but are time-consuming, expensive and require antibodies for each transcription factor \citep{Elnitski2006}. Accurate computational methods to infer gene regulatory networks, particularly methods that leverage genome-scale experimental data, are urgently required not only to supplement empirical approaches but also, if possible, to explore these data in new, more-integrative ways.

Many computational methods have been developed to infer regulatory networks from gene expression data, predominately employing unsupervised techniques. Several comparisons have been made of network inference methods, but a comprehensive evaluation that covers unsupervised, semi-supervised and supervised methods is lacking, and many questions remain open.
Here we address fundamental questions, including which methods are suitable for what kinds of experimental data types, and how many samples these methods require.

The most-recent and largest comparison so far has been performed by \citet{Madhamshettiwar2012}. They compared the prediction accuracy of eight unsupervised and one supervised method on 38 simulated data sets. The methods showed large differences in prediction accuracy but the supervised method was found to perform best, despite the parameters of the unsupervised methods having been optimized. Here we extend this study to 17 unsupervised methods and include a direct comparison with supervised and semi-supervised methods on a wide range of networks and experimental data types (knock-out, knock-down and multi-factorial).

Another comprehensive evaluation, limited to unsupervised methods, has been performed as part of the Dialogue for Reverse Engineering Assessments and Methods (DREAM), an annual open competition in network inference \citep{Stolovitzky2007,Stolovitzky2009,Marbach2010,Prill2010,Marbach2012}. Results from DREAM highlight that network inference is a challenging problem. To quote~\citet{Prill2010}:  ``The vast majority of the teams' predictions were statistically equivalent to random guesses.'' However, an important result of the DREAM competition is that under certain conditions simple methods can perform well: ``...the z-score prediction would have placed second, first, and first (tie) in the 10-node, 50-node, and 100-node subchallenges, respectively''~\citep{Prill2010}.

Unsupervised methods rely on expression data only but tend to achieve lower prediction accuracies than supervised methods \citep{Mordelet2008,Cerulo2010,Madhamshettiwar2012}. By contrast, supervised methods require information about known interactions for training, and this information is typically sparse. Semi-supervised methods reflect a compromise and can be trained with much fewer interaction data, but usually are not as accurate predictors as supervised methods. One of the few comparisons with supervised methods was performed by \citet{Mordelet2008}. They evaluated SIRENE (Supervised Inference of Regulatory Networks) in comparison to CLR, ARACNE, Relevance Networks (RN) and a Bayesian Network on an {\it E. coli} benchmark data set by \citet{Faith2007} and found that the supervised method considerably outperformed the unsupervised techniques.

\citet{Cerulo2010} compared supervised and semi-supervised support vector machines with two unsupervised methods and found the former superior. Our evaluation employs similar supervised and semi-supervised methods but includes many more unsupervised methods, distinguishes between experimental types and performs replicates, resulting in a more-complete picture. A related evaluation by \citet{Schaffter2011} compared six unsupervised methods on larger networks with 100, 200 and 500 nodes and simulated expression data. Again the z-score method was found to be one of the top performers in knock-out experiments.  

Several smaller evaluations have been performed but are largely restricted to four unsupervised methods (ARACNE, CLR, MRNET and RN) in comparisons with a novel approach on small data sets. The ARACNE method was introduced by \citet{Margolin2006} and showed superior precision and recall when compared to RN and a Bayesian Network algorithm on simulated networks. 
\citet{Meyer2007} compared all four unsupervised inference algorithms on large yeast sub-networks (100 up to 1000 nodes) using simulated expression data, and \citet{Altay2010} investigated the bias in the predictions of those algorithms. \citet{Faith2007} evaluated CLR, ARACNE, RN and a linear regression model on {\it E. coli} interaction data from RegulonDB and found CLR to outperform the other methods. \citet{Lopes2009} studied the prediction accuracy of ARACNE, MRNET, CLR and SFFS+MCE, a feature selection algorithm, on simulated networks and found the latter superior for networks with small node degree.
\citet{Haynes2009} developed a synthetic regulatory network generator (GRENDEL) and measured the prediction accuracy of ARACNE, CLR, DBmcmc and {Symmetric-N} for various network sizes and different experimental types.
\citet{Werhli2006} compared RN, graphical Gaussian models (GGMs) and Bayesian networks (BNs) on the Raf pathway, a small cellular signalling network with 11 proteins, and on simulated data. BNs and GGMs were found to outperform RN on observational data.
\citet{Camacho2007} compared Regulatory Strengths Analysis (RSA), Reverse Engineering by Multiple Regression (NIR), Partial Correlations (PC) and Dynamic Bayesian Networks (BANJO) on a small, simulated network with 10 genes, with different levels of noise. In the noise-free scenario the PC method showed the highest accuracy. Finally, \citet{Cantone2009} constructed a small, synthetic, {\it in vivo} network of five genes and measured time series and steady-state expression. In an evaluation of BANJO, ARACNE and two models based on ordinary differential equations they found the latter two to achieve the highest accuracies. \citet{Bansal2007} also evaluated BANJO, ARACNE and ordinary differential equations but on random networks and simulated expression data.

In the following sections we first describe the different inference methods in detail, before evaluating their prediction accuracies on simulated gene expression data and regulatory networks of varying size. We continue with a discussion of the prediction results and conclude with guidelines for the use of the evaluated methods.

\section{Methods}

We compared the prediction performance of unsupervised, semi-supervised and supervised network inference methods. Following other authors \citep{Husmeier2003,Mordelet2008,Haynes2009} we assess prediction performance by the Area under the Receiver Operator Characteristic curve (AUC)
\begin{equation}
  AUC = \frac{1}{2} \sum_{k=1}^n (X_k - X_{k-1})(Y_k + Y_{k-1}),
\end{equation}
where $X_k$ is the false positive rate and $Y_k$ is the true positive rate for the $k$-th output in the ranked list of predicted edge weights. An AUC of 1.0 indicates perfect prediction, while an AUC of 0.5 indicates a performance no better than random guessing. 

Note that in contrast to other measures such as F1 score, Matthews correlation, recall or precision  \citep{Baldi2000}, AUC does not require choice of a threshold to infer interactions from predicted weights; rather, it compares the predicted weights directly to the topology of the true network. In the Supplementary Material we nonetheless report results based F1 score and Matthews correlation.

To avoid discrepancies between the gene expression values generated by true networks and the actually known, partial networks, we performed our evaluations on simulated, steady-state expression data, generated from sub-networks extracted from {\it E. coli} and {\it S. cerevisiae} networks. This allowed us to assess the accuracy of an algorithm against a perfectly known true network \citep{Bansal2007}. 
When comparing the true with the inferred network, the direction and type of interactions were ignored, since many inference methods can infer only the existence of an interaction. For the same reason self-interactions were excluded from the network comparison.
We employed \emph{GeneNetWeaver} \citep{Marbach2009,Schaffter2011} and \emph{SynTReN} \citep{VandenBulcke2006} to extract sub-networks and to simulate gene expression data.

\emph{GeneNetWeaver} has been part of several evaluations, most prominently the DREAM challenges. The simulator extracts sub-networks from known interaction networks such as those of {\it E. coli} and {\it S. cerevisiae}, emulates transcription and translation, and employs a set of ordinary differential equations describing chemical kinetics to generate expression data for knock-out, knock-down and multi-factorial experiments. 

To simulate knock-out experiments the expression value of each gene is in turn set to zero, whereas for knock-down experiments the expression value is halved. In multi-factorial experiments the expression levels of a small number of genes is perturbed by a small, random amount. 

\emph{SynTReN} is a similar but older simulator. Sub-graphs are also extracted from {\it E. coli} and {\it S. cerevisiae} networks but it simulates only the transcription level and multi-factorial experiments. However, \emph{SynTReN} is faster than \emph{GeneNetWeaver} and allows one to vary the sample number independently of the network size. 

To enable a comprehensive and fair comparison we evaluated the prediction accuracies of these inference methods on sub-networks with different numbers of nodes (10,...,110) extracted from {\it E. coli} and {\it S. cerevisiae}, and used three experimental data types (knock-out, knock-down, multi-factorial) with varying sample set sizes (10,...,110)) simulated by GeneNetWeaver and SynTReN. 

We performed no parameter optimization for unsupervised methods, since this would require training data (known interactions) and render those methods supervised. For the supervised and semi-supervised methods, 5-fold cross-validation was applied and parameters were optimized on the training data only. The following sections describe the inference methods in detail.

\subsection{Unsupervised}

This section describes the evaluated unsupervised methods. CLR, ARACNE, MRNET and MRNET-B are part of the R package ``minet'' and were called with their default parameters~\citep{Meyer2008}, with the exception of ARACNE. With the default parameter $eps=0.0$, ARACNE performed very poorly and we used $eps=0.2$ instead. Similarly, GENIE~\citep{Huynh-Thu2010}, MINE~\citep{Reshef2011}, and PCIT~\citep{Reverter2008} were installed and evaluated with default parameters. All other methods were implemented according to their respective publications. SPEARMAN-C, EUCLID and SIGMOID are implementations of our own inference algorithms.

\subsubsection{Correlation}
-based network inference methods assume that correlated expression levels between two genes are indicative of a regulatory interaction. Correlation coefficients range from +1 to -1 and a positive correlation coefficient indicates an activating interaction, while a negative coefficient indicates an inhibitory interaction. The common correlation measure by Pearson is defined as

\begin{equation}
  corr(X_i, X_j) = \frac{cov(X_i, X_j)}{\sigma(X_i) \cdot \sigma(X_j)},
\end{equation}
where $X_i$ and $X_j$ are the expression levels of genes $i$ and $j$, $cov( \cdot,\cdot )$ denotes the covariance, and $\sigma(\cdot)$ is the standard deviation. Pearson's correlation measure assumes normally distributed values, an assumption that does not necessarily hold for gene-expression data. Therefore rank-based measures are frequently employed, with the measures by Spearman and Kendall being the most common. Spearman's method is simply Pearson's correlation coefficient for the ranked expression values, and Kendall's $\tau$ coefficient is computed as

\begin{equation}
  \tau(X_i, X_j) = \frac{con(X_i^r, X_j^r) - dis(X_i^r, X_j^r)}{\frac{1}{2} n (n-1)},
\end{equation}
where $X_i^r$ and $X_j^r$ are the ranked expression profiles of genes $i$ and $j$. $con( \cdot,\cdot )$ denotes the number of concordant and $dis( \cdot,\cdot )$ the number of disconcordant value pairs in $X_i^r$ and $X_j^r$, with both profiles being of length $n$.

Since our evaluation of prediction accuracy does not distinguish between inhibiting and activating interactions, the predicted interaction weights are computed as the absolute value of the correlation coefficients
\begin{equation}
  w_{ij} = |corr(X_i, X_j)|.
\end{equation}

\subsubsection{SPEARMAN-C}
is a modification of Spearman's correlation coefficient where we attempted to favor hub nodes, which have many, strong interactions. The correlation coefficient is corrected by multiplying it by the mean correlation of gene $i$ with all other genes $k$, and the absolute value is taken as the interaction weight

\begin{equation}
	 w_{ij} = | corr(X_i, X_j) \cdot \frac{1}{n}\sum_k^n corr(X_i, X_k) | ,
\end{equation}
where $corr(\cdot,\cdot)$ is Spearman's correlation coefficient.

\subsubsection{WGCNA}
stands for Weighted Gene Co-expression Network Analysis~\citep{Langfelder2008} and is a modification of correlation-based inference methods that amplifies high correlation coefficients by raising the absolute value to the power of $\beta$ (``softpower'')
	 
\begin{equation}
  w_{ij} = |corr(X_i, X_j)|^{\beta}, 
\end{equation}
with $\beta \geq 1$. Since softpower is a non-linear but monotonic transformation of the correlation coefficient, the prediction accuracy measured by AUC will be no different from that of the underlying correlation method itself. Consequently we show only results for correlation methods but not for the WGCNA modification, which would be identical.

\subsubsection{RN}
(relevance networks) by \citet{Butte2000} measure the mutual information (MI) between gene expression profiles to infer interactions. The mutual information $I$ between discrete variables $X_i$ and $X_j$ is defined as

\begin{equation}
  I(X_i, X_j) = \sum_{x_i \in X_i} \sum_{x_j \in X_j} p(x_i, x_j) 
                     \log \left ( \frac{p(x_i, x_j)}{p(x_i)p(x_j)} \right ),
\end{equation}
where $p(x_i, x_j)$ is the joint probability distribution of $X_i$ and $X_j$, and $p(x_i)$ and $p(x_j)$ are the marginal probabilities. $X_i$ and $X_j$ are required to be discrete variables. We used equal-width binning for discretization and empirical entropy estimation as described by~\citet{Meyer2008}.

\subsubsection{CLR} 
is the abbreviation for Context Likelihood of Relatedness \citep{Faith2007} and  extends the relevance network method (RN) by taking the background distribution of the mutual information values $I(X_i, X_j)$ into account. The most probable interactions are those that deviate most from the background distribution and for each gene $i$ a maximum z-score $z_i$ is calculated as 
	 
\begin{equation}
  z_i = \max_j \left (  0, \frac{I(X_i, X_j)- \mu_i}{\sigma_i}  \right ),
\end{equation}
where $\mu_i$ and $\sigma_i$ are the mean value and standard deviation, respectively, of the mutual information values $I(X_i, X_k$), $k=1,...,n$. The interaction $w_{ij}$ between two genes $i$ and $j$ is then defined as

\begin{equation}
  w_{ij} = \sqrt{z_i^2+z_j^2} .
\end{equation}
The background correction step aims to reduce the prediction of false interactions based on spurious correlations and indirect interactions.

\subsubsection{ARACNE} 
stands for Algorithm for the Reconstruction of Accurate Cellular Networks \citep{Margolin2006}, and is another modification of the relevance network that applies the Data Processing Inequality (DPI) to filter out indirect interactions. The DPI states that, if gene $i$ interacts with gene $j$ via gene $k$, then the following inequality holds:

\begin{equation}
  I(X_i, X_j) \leq \min( I(X_i, X_j), I(X_j, X_k) ).
\end{equation}
	 
ARACNE considers all possible triplets of genes (interaction triangles) and computes the mutual information values for each gene pair within the triplet. Interactions within an interaction triangle are assumed to be indirect and are therefore pruned if they violate the DPI beyond a specified tolerance threshold $eps$. We used an threshold of $eps=0.2$ for our evaluations.

\subsubsection{PCIT} 
is an abbreviation of Partial Correlation and Information Theory \citep{Reverter2008} and is similar to ARACNE. PCIT  extracts all possible interaction triangles and applies the DPI to filter indirect interactions, but uses partial correlation coefficients instead of mutual information as interaction weights. The partial correlation coefficient $corr_{ij}^\text{partial}$ between two genes $i$ and $j$ within an interaction triangle $(i,j,k)$ is defined as

\begin{equation}
  corr_{ij}^\text{partial} = \frac{ corr(X_i, X_j) - corr(X_i, X_k)corr(X_j, X_k) }
                               { \sqrt{(1-corr(X_i, X_k))^2 (1-corr(X_j, X_k))^2} },
\end{equation}
where $corr(\cdot,\cdot)$ is Person's correlation coefficient. The partial correlation coefficient aims to eliminate the effect of the third gene $k$ on the correlation of genes $i$ and $j$.

\subsubsection{MRNET}
\citep{Meyer2007} employs mutual information between expression profiles and a feature selection algorithm (MRMR) to infer interactions between genes. More precisely, the method places each gene in the role of a target gene $j$ with all other genes $V$ as its regulators. The mutual information between the target gene and the regulators is calculated and the Minimum-Redundancy-Maximum-Relevance (MRMR) method is applied to select the best subset of regulators. MRMR step-by-step builds a set $S$ by selecting the genes $i^{MRMR}$ with the largest mutual information value and the smallest redundancy based on the following definition

\begin{equation}
  i^{MRMR} = \underset{i \in V \setminus S}{\operatorname{argmax}} (s_i),
\end{equation}

with $s_i = u_i - r_i$. The relevance term $u_i = I(X_i, X_j)$ is thereby the mutual information between gene $i$ and target $j$, and the redundancy term $r_i$ is defined as 
	 
\begin{equation}
  r_i = \frac{1}{|S|} \sum_{k \in S} I(X_i, X_k).
\end{equation}

Interaction weights $w_{ij}$ are finally computed as $w_{ij} = \max (s_i, s_j)$.

\subsubsection{MRNET-B}
is a modification of MRNET that replaces the forward selection strategy to identify the best subset of regulator genes by a backward selection strategy followed by a sequential replacement~\citep{Meyer2010}.

\subsubsection{GENIE}
(GEne Network Inference with Ensemble of trees)
is similar to MRNET in that it also lets each gene take on the role of a target regulated by the remaining genes and then employs a feature selection procedure to identify the best subset of regulator genes. In contrast to MRNET, Random Forests and Extra-Trees are used for regression and feature selection~\citep{Huynh-Thu2010} rather than mutual information and MRMR.

\subsubsection{SIGMOID}
models the regulation of a gene by a linear combination with
soft thresholding. The predicted expression value $X_{ik}^\prime$ of gene $i$ at time point $k$ is described by the sum over the weighted expression values $X_{jk}$ of the remaining genes, constrained by a sigmoid function $\sigma(\cdot)$

\begin{align}
  & X_{ik}^\prime = \sigma( \sum_{j \neq i}^n X_{jk} w_{ij} + b_i) \\
  & \sigma(x) = \frac{1}{1+e^{-x}}. 
\end{align}
The regulatory weights $w_{ij}$ are determined by minimizing the following quadratic error function over the predicted expression values $X_{ik}^\prime$ and the observed values $X_{ik}$:

\begin{equation}
  E(w,b) = \frac{1}{2} \sum_i \sum_k ( X_{ik}^\prime - X_{ik} )^2 .
\end{equation} 
Finally, the interaction weights $w_{ij}^\prime$ for the undirected network are computed by averaging over the forward and backward weights:

\begin{equation}
  w_{ij}^\prime = \frac{|w_{ij}|+|w_{ji}|}{2}.
\end{equation}

\subsubsection{MD}
(Mass-Distance) by \citet{Yona2006} is a similarity measure for expression profiles. It estimates the probability to observe a profile inside the volume delimited by the profiles. The smaller the volume, the more similar are the two profiles. Given two expression profiles $X_i$ and $X_j$, the total probability mass of samples whose $k$-th feature is bounded between the expression values $X_{ik}$ and $X_{jk}$ is calculated as  
    
\begin{equation}
  \text{MASS}_k(X_i, X_j) = \sum_{min(X_{ik}, X_{jk}) \leq x \leq max(X_{ik}, X_{jk})} freq(x),
\end{equation} 
with $freq(x)$ is the empirical frequency. The mass distance $\text{MD}_{ij}$ is defined as the total volume of profiles bounded between the two expression profiles $X_i$ and $X_j$ and is estimated by the product over all coordinates $k$
\begin{equation}
  \text{MD}_{ij} = \prod_{k}^{n} \text{MASS}_k(X_i, X_j),
\end{equation} 
with $n$ is the length of the expression profiles. Since the $\text{MD}_{ij}$ is symmetric and positive we interpret it directly as an interaction weight $w_{ij}$.

\subsubsection{MR}
(mutual rank) by \citet{Obayashi2009} employs ranked Pearson's correlation as a measure to describe gene coexpression. For a gene $i$, first Pearson's correlation with all other genes $k$ 
is computed and ranked. Then the rank achieved for gene $j$ is taken as score to describe the similarity of the gene expression profiles $X_i$ and $X_j$:
   \begin{equation}
  rank_{ij} = \underset{j}{\operatorname{rank}}  ( corr(X_i,X_k), \ \forall k  \neq i ),
\end{equation}  
with $corr(\cdot, \cdot)$ being Pearson's correlation coefficient. The final interaction weight $w_{ij}$ is calculated as the geometric average of the ranked correlation between gene $i$ and $j$ and vice versa:
\begin{equation}
  w_{ij} = \frac{rank_{ij} \cdot rank_{ji}}{2}.
\end{equation}

\subsubsection{MINE}
is a class of Maximal Information-based Nonparametric Exploration statistics by \citet{Reshef2011}. The Maximal Information Coefficient (MIC) is part of this class and a novel measure to quantify non-linear relationships. We computed the $\text{MIC}$ for expression profiles $X_i$ and $X_j$ and interpreted the $\text{MIC}$ score as an interaction weight

\begin{equation}
  w_{ij} = \text{MIC}(X_i,X_j).
\end{equation}

\subsubsection{EUCLID}
is a simple method that employs the euclidean distance between the normalized expression profiles $X_i^\prime$ and $X_j^\prime$ of two genes as interaction weights
 
\begin{equation}
  w_{ij} = \sqrt{ \sum_k (X_{ik}^\prime-X_{jk}^\prime)^2 } ,
\end{equation}  
where profiles are normalized by computing the absolute difference of expression values $X_{ik}$ to the median expression in profile $X_i$ 
\begin{equation}
  X_{ik}^\prime = |X_{ik} - \text{median}(X_i)|.
\end{equation}

\subsubsection{Z-SCORE}
is a network inference strategy by \citet{Prill2010} that takes advantage of knock-out data. It assumes that a knock-out affects directly interacting genes more strongly than others. The z-score $z_{ij}$ describes the effect of a knock-out of gene $i$ in the $k$-th experiment on gene $j$ as the normalized deviation of the expression level $X_{jk}$ of gene $j$ for experiment $k$ from the average expression $\mu(X_j)$ of gene $j$:

\begin{equation}
	 z_{ij} = |\frac{X_{jk}-\mu(X_j)}{\sigma(X_j)}|.
	 \label{eq:z-score}
\end{equation}
The original Z-SCORE methods requires knowledge of the knock-out experiment $k$ and is therefore not directly applicable to data from multi-factorial experiments. The method, however, can easily be generalized by assuming that the minimum expression value within a profile indicates the knock-out experiment ($\min(X_{j}) = X_{jk}$). Equation~\ref{eq:z-score} then becomes

\begin{equation}
	 w_{ij} = |\frac{\min(X_j)-\mu(X_j)}{\sigma(X_j)}|,
\end{equation}
and the method can be applied to knock-out, knock-down and multi-factorial data. Note that $z_{ij}$ is an asymmetric score and we therefore take the maximum of $z_{ij}$ and $z_{ji}$ to compute the final interaction weight $w_{ij}$ as

\begin{equation}
	 w_{ij} = \max( z_{ij}, z_{ji} ).
\end{equation}

\subsection{Supervised} 

A great variety of different supervised machine learning methods has been developed. We limit our evaluation to Support Vector Machines (SVMs) because they have been successfully applied to the inference of gene regulatory networks \citep{Mordelet2008} and can easily be trained in a semi-supervised setting \citep{Cerulo2010}. We used the SVM implementation \emph{SVMLight} by \citet{Joachims1999} for all evaluations.

SVMs are trained by maximizing a constrained, quadratic optimization problem over Lagrange multipliers $\alpha$:
\begin{equation}
  \newcommand{\ve}[1] {\mathbf{#1}}
  \newcommand{\sprod}[2] {\ve{#1}^T\ve{#2}}
  \newcommand{\maximum}[1] {\underset{#1}{\mathop{\rm max}}\:}  
  \begin{split}
   & { \maximum{\ve{\alpha}}{\ L(\ve{\alpha})} = \sum_{i=1}^N \alpha_i -
              \frac{1}{2} \sum_{i,j=1}^N \alpha_i \alpha_j\, y_i y_j\,
              {\ve{x}_i}^T \ve{x}_j } \\
   & { \text{subject to }
       \begin{cases}
         \,\sum_{i=1}^N \alpha_i y_i = 0  \\
         \,0 \leq \alpha_i \leq C \ \ \text{for}\ \forall i.
       \end{cases}
     }
  \end{split}
\end{equation}

The labels $y_i$ determine the class to which feature vector $\mathbf{x}_i$ belongs and $C$ is the so-called \emph{complexity} parameter that needs to be tuned for optimal prediction performance. Once the optimal Lagrange multipliers $\alpha$ are found, a feature vector can be classified by its 
signed distance $d(\mathbf{x})$ to the decision boundary, which is computed as

\begin{equation}
  \newcommand{\ve}[1] {\mathbf{#1}}
  d(\ve{x}) = \sum_{i=1}^N \alpha_i y_i \, {\ve{x}_i}^T \ve{x} + b.
\end{equation}
The distance $d(\mathbf{x})$ can be interpreted as a confidence value. The larger the absolute distance, the more confident the prediction, and similar to a correlation value we interpret the distance as an interaction weight. 

In contrast to unsupervised methods, e.g. correlation methods, the supervised approach does not directly operate on pairs of expression profiles but on feature vectors that can be constructed in various ways. We computed the outer product of two gene expression profiles $X_i$ and $X_j$ to construct feature vectors:

\begin{equation}
  \newcommand{\ve}[1] {\mathbf{#1}}
  \ve{x} = X_i X_j^T .
\end{equation}
The outer product was chosen because it is commutative, and predicted interactions are therefore symmetric and undirected. A sample set for the training of the SVM is then composed of feature vectors $\mathbf{x}_i$ that are labeled $y_i = +1$ for gene pairs that interact and $y_i = -1$ for those that do not interact. 

\begin{figure}[ht] 
\centerline{\includegraphics[scale=0.70]{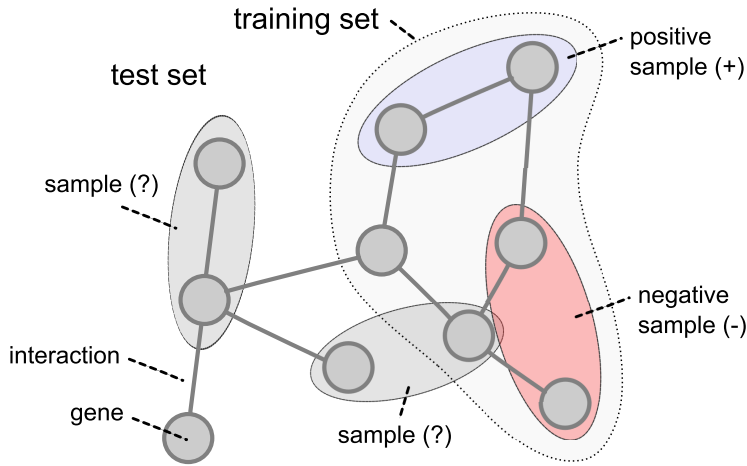}}
\caption{Extraction of samples for the training and test set from a gene interaction network.}
\label{fig:Network} 
\end{figure}

If all gene pairs are labeled, all network interactions would be known and prediction would be unnecessary. In practice and for evaluation purposes training is therefore performed on a set of labeled samples, and predictions are generated for the samples of a test set. Figure~\ref{fig:Network} depicts the concept. All samples  within the training set are labeled and all remaining gene pairs serve as test samples.

Note that the term ``sample'' in the context of supervised learning refers to a feature vector derived from a pair of genes and their expression profiles, whereas a sample in an expression data set refers to the gene expression values for a single experiment, e.g. a gene knock-out. 

We evaluate the prediction accuracy of the supervised method by generating labeled feature vectors for all gene pairs (samples) of a network. This entire sample set is then divided in to five parts. Each of the parts is used as a test set and the remaining four parts serve as a training set. The total prediction accuracy is averaged over the prediction accuracies achieved during the five iterations (five-fold cross-validation).

\subsection{Semi-supervised} 

Data describing regulatory networks are sparse and typically only a small fraction of the true interactions is known. The situation is even worse for negative data (non-interactions), since experimental validation largely aims to detect but not exclude interactions. The case that all samples within a training data set can be labeled as positive or negative is therefore rarely given for practical network inference problems and supervised methods are limited to very small training data sets, which negatively affects their performance.

\begin{figure}[ht] 
\centerline{\includegraphics[scale=1.3]{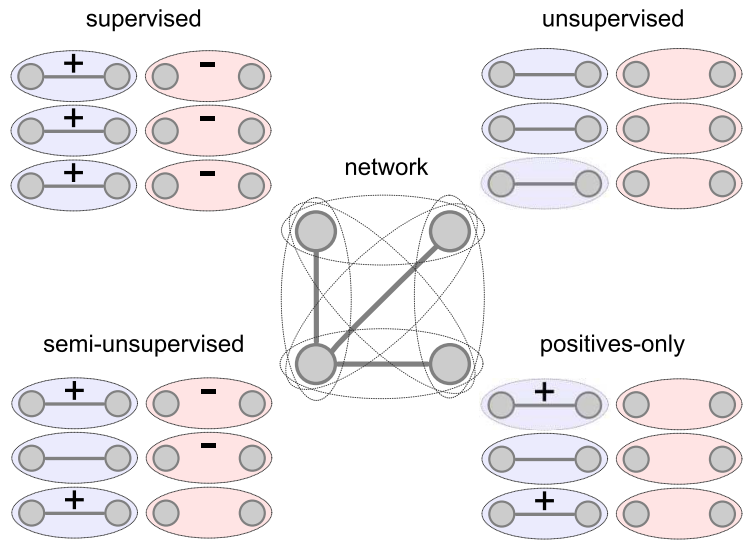}}
\caption{Original labeling of samples for supervised, unsupervised, semi-supervised and positives-only prediction methods. All the six samples within a sample set are generated by a four-node network with three interactions.}
\label{fig:Labeling} 
\end{figure}

Semi-supervised methods strive to take advantage of the unlabeled samples within a training set by taking the distribution of unlabeled samples into account, and can even be trained on positively labeled data only. Figure~\ref{fig:Labeling} shows the required labeling of data for the different approaches. Supervised methods require all samples within the training set to be labeled, while unsupervised methods require no labeling at all. Semi-supervised approaches can be distinguished into methods that need positive and negative samples and methods that operate on positive samples only. 

The semi-supervised method used in this evaluation is based on the supervised SVM approach described above. The only difference is in the labeling of the training set. In the semi-supervised setting only a portion of the training samples is labeled. To enable the SVM training, which requires all samples to be labeled, all unlabeled samples within the semi-supervised training data are relabeled as negatives \citep{Cerulo2010}. This approach enables a direct comparison of the same prediction algorithm trained with fully or partially labeled data.

We assigned different percentages (10\%...,100\%) of true positive and negative or positive-only labels to the training set. The prediction performance of the different approaches was then evaluated by five-fold cross-validation, with equal training/test set sizes for the supervised, semi-supervised, positives-only and unsupervised methods compared.

\section{Results}

In the following we first evaluate the prediction accuracy of unsupervised methods before comparing
two selected unsupervised methods with supervised and semi-supervised approaches.

\subsection{Unsupervised methods} 

\begin{figure}[ht] 
\centerline{\includegraphics[scale=0.6]{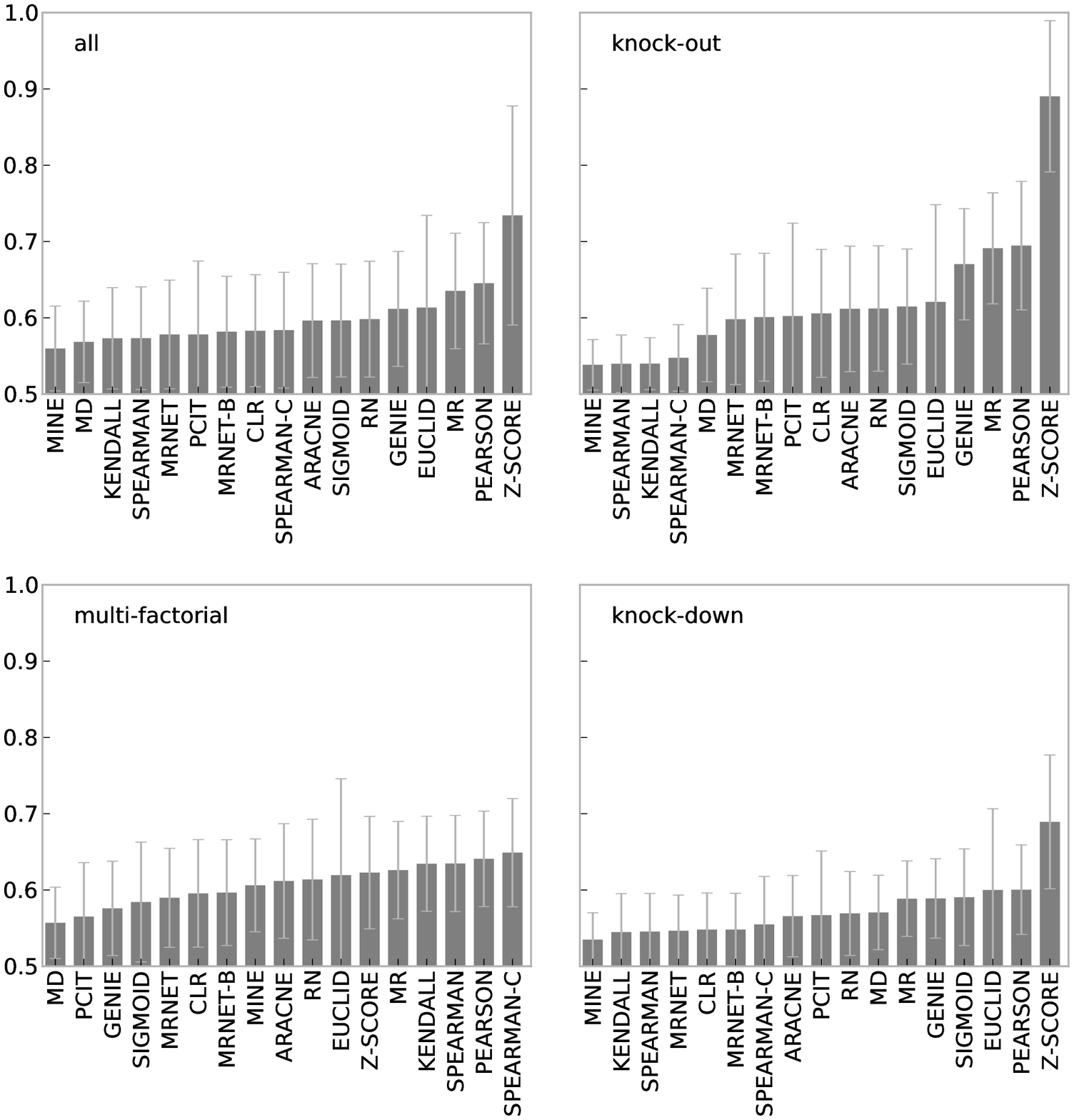}}
\caption{Prediction accuracy (AUC) of unsupervised methods on multi-factorial, knock-out, knock-down and averaged (all) data generated by GenNetWeaver. 10 repeats over networks with 10,...,110 nodes, extracted from {\it E. coli} and {\it S. cerevisiae}. Error bars show standard deviation.}
\label{fig:Unsupervised} 
\end{figure}

Figure~\ref{fig:Unsupervised} shows the prediction accuracies measured by AUC for all unsupervised methods for three different experimental types (knock-out, knock-down and multi-factorial) and the average AUC (all) over the three types.
Networks with 10, 30, 50, 70, 90 and 110 nodes were extracted from {\it E. coli} and {\it S. cerevisiae} and expression data were simulated with GenNetWeaver, with the number of samples (experiments) equal to the nodes of the network. Every evaluation was repeated 10 times, so each bar therefore represents an AUC averaged over 60 networks or 180 networks (all).  

Most obvious are the large standard deviations in prediction accuracy across all methods and experimental types. For small networks the accuracy of a method can easily vary between no better than guessing to close to perfect (see Supplementary Material). While most differences between methods are statistically significant (p-values $<$ 0.01 for Wilcoxon rank sum test with Bonferroni correction), differences are largely small and the ranking for most methods is therefore not stable and depends on the experimental data type, the source network, the sub-network size and other factors (see Supplementary Material). However, a simple Pearson's correlation is consistently the second-best performer for all experimental types.

Interestingly, rank-based correlation methods (SPEARMAN, KENDALL) that are very similar to Pearson correlation perform very poorly on knock-out and knock-down data but well for multi-factorial experiments.

With the exception of the Z-SCORE method prediction, accuracies are very low in general. Z-SCORE was specifically designed for knock-out data and indeed clearly outperforms all other methods for this experimental type, despite its simplicity. It is the only unsupervised method that achieves a good prediction accuracy (AUC = 0.9).

\subsection{Network size} 

\begin{figure}[ht] 
\centerline{\includegraphics[scale=0.7]{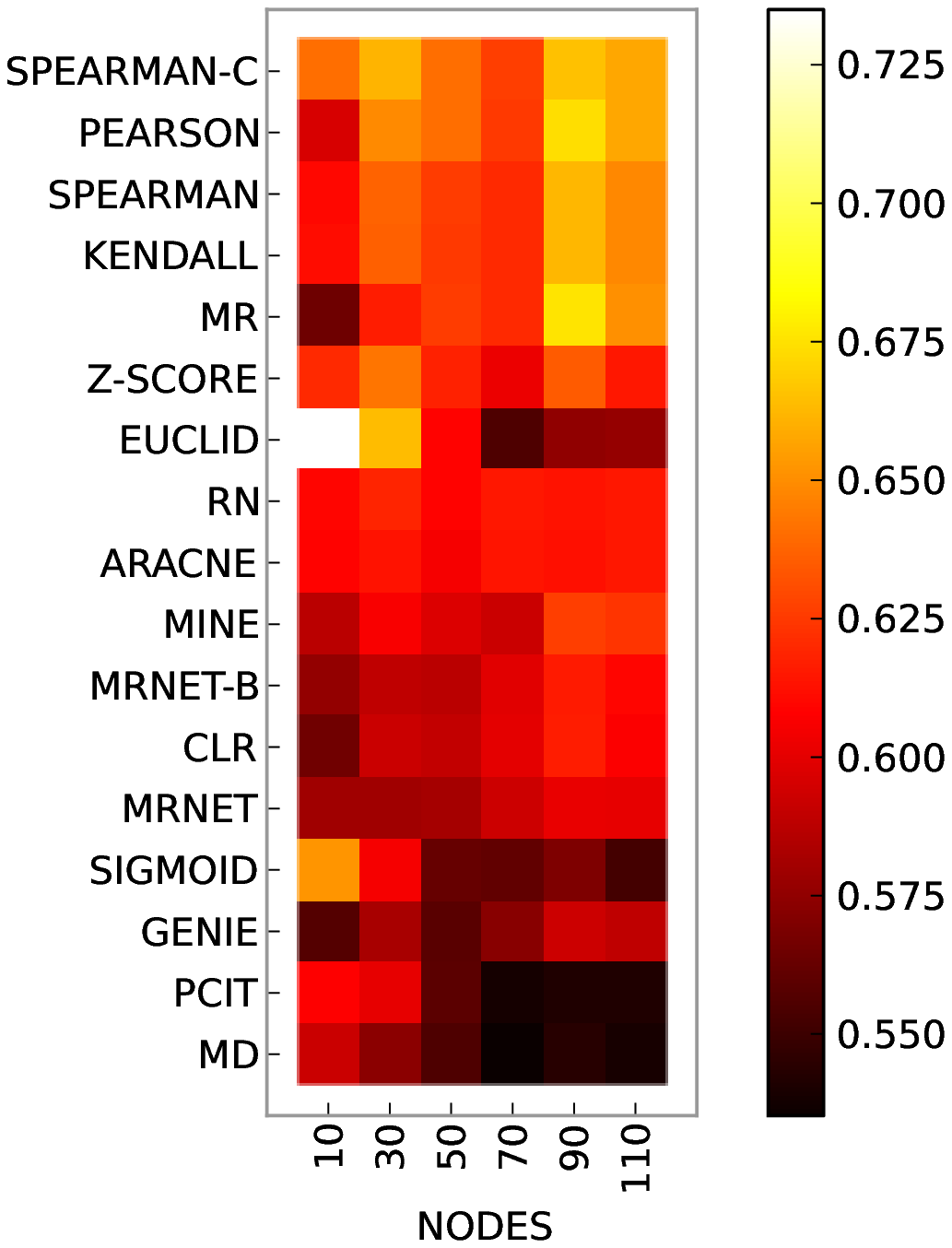}}
\caption{Prediction accuracy (AUC) of unsupervised methods on multi-factorial data for different network sizes. Data generated by GenNetWeaver and extracted from {\it E. coli} and {\it S. cerevisiae}.}
\label{fig:UnsupervisedHeatmap} 
\end{figure}

Figure~\ref{fig:Unsupervised} summarizes results averaged over networks. We also examined how the network size impacts the prediction performance of the various methods. The heat map in Figure~\ref{fig:UnsupervisedHeatmap} is based on the same data as Figure~\ref{fig:Unsupervised}, but shows the prediction accuracies (AUC) of the inference methods on multi-factorial data for networks with different numbers of nodes (see Supplementary Material for the related figures on knock-out and knock-down data).

The rows in Figure~\ref{fig:UnsupervisedHeatmap} are ordered according to mean AUC and the ranking is therefore identical to that in the multi-factorial bar graph in Figure~\ref{fig:Unsupervised}. Top performers on average are the correlation methods by Pearson, Spearman and Kendall, with the corrected Spearman method (SPEARMAN-C) achieving the highest mean AUC. However, when focusing on networks of specific size, the best performance is achieved by the EUCLID method for small networks with 10 nodes. Other methods also show different behaviors with respect to network size. Correlation methods clearly achieve higher AUCs for large networks. Similar trends can be observed for MR, MINE, GENIE, MRNET, MRNET-B and CLR. In contrast, SIGMOID, PCIT and MD decrease in prediction accuracy for growing network sizes, while the performance of RN and ARACNE is seemingly unaffected by network size within the investigated size range.

\subsection{Sample number} 

\begin{figure}[ht] 
\centerline{\includegraphics[scale=0.6]{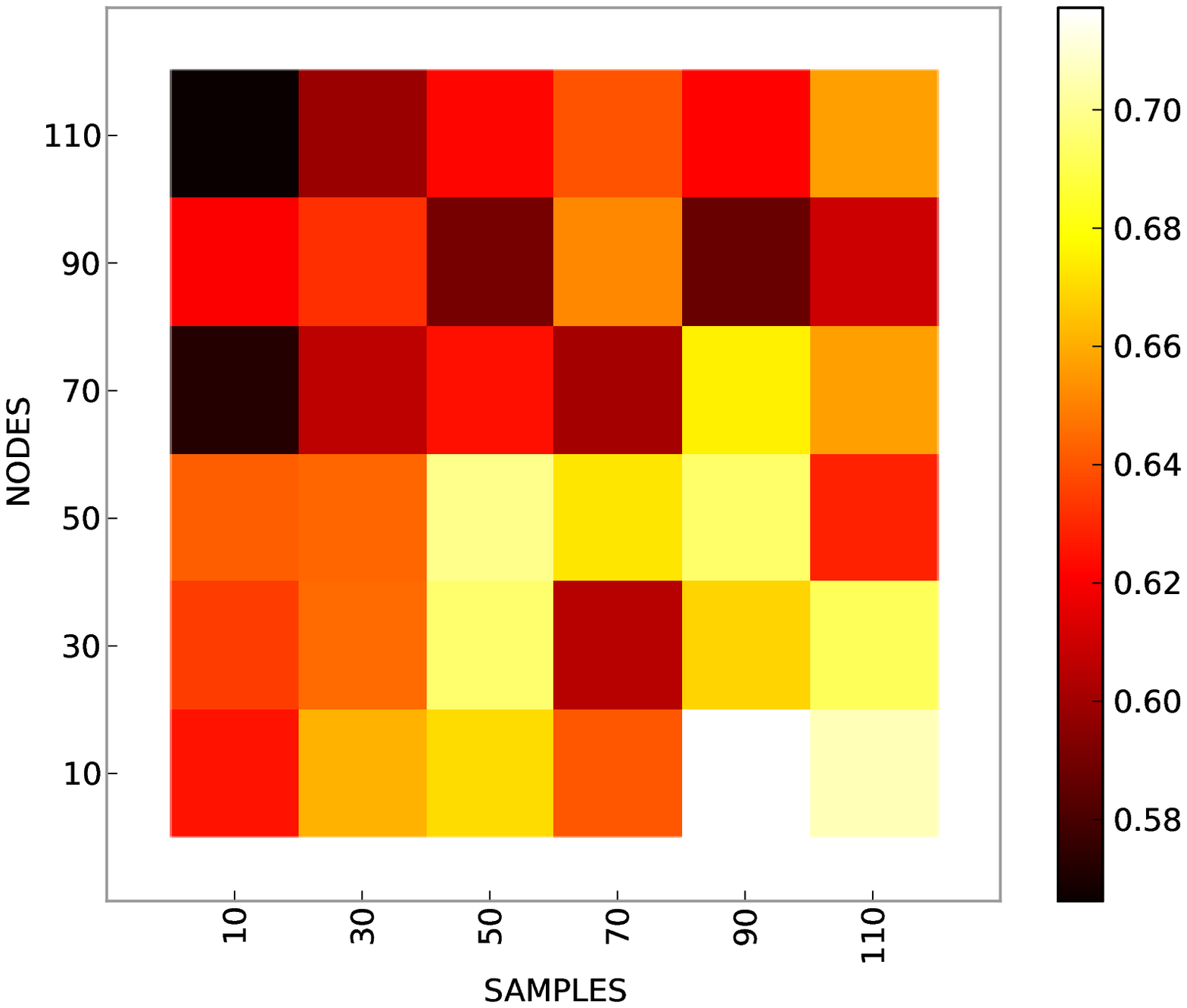}}
\caption{Prediction accuracy (AUC) averaged over all unsupervised methods on multi-factorial for different network sizes (nodes) and sample numbers. Data generated by SynTReN and extracted from {\it E. coli}. 10 repeats. }
\label{fig:UnsupervisedSamples} 
\end{figure}

Apart from the size of the network, we also expected the number of samples to have an effect on the prediction accuracy of the inference algorithms. GenNetWeaver generates gene expression profiles with the same number of samples as network nodes (genes). We therefore used SynTReN to vary networks size and sample number independently. The heat map in Figure~\ref{fig:UnsupervisedSamples} shows prediction accuracy (AUC) averaged over all inference methods for different network sizes and sample numbers. SynTReN simulates expression data for multi-factorial experiments only, and networks were extracted from {\it E. coli}. All experiments were repeated 10 times. The results show the expected trend of improving accuracy with increasing number of samples and decreasing size of network.

However, the absolute improvements in prediction accuracy are rather small with additional data, most likely because unsupervised methods can infer only simple network topologies reliably and small sample sets are sufficient for this purpose. For instance, networks with 50 nodes are predicted with an AUC of roughly 0.65, when 50 samples are available. Increasing the sample set size to 110 raises the prediction accuracy only to an AUC of around 0.67.

\subsection{Supervised methods} 

Finally, we wanted to compare unsupervised with supervised and semi-supervised approaches. Because of the time-consuming training required for supervised methods we limited our evaluation to networks with 30 nodes extracted from {\it E. coli} networks. Expression profiles were generated with GenNetWeaver, and each experiment was repeated 10 times. 

\begin{figure}[ht] 
\centerline{\includegraphics[scale=0.6]{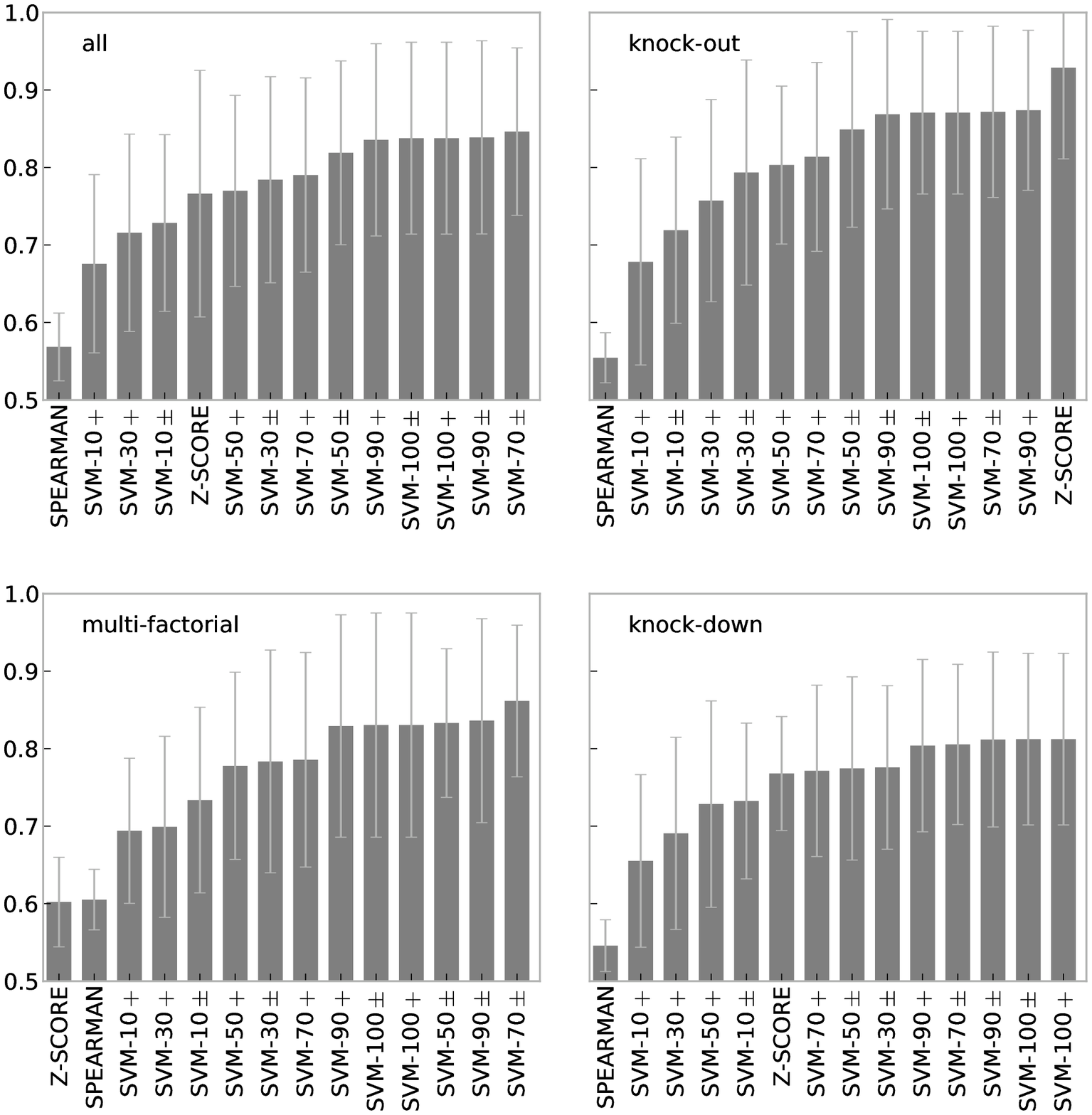}}
\caption{Prediction accuracy (AUC) of supervised methods on multi-factorial, knock-out, knock-down and averaged (all) data generated by GenNetWeaver. five-fold cross-validation and 10 repeats over networks with 30 nodes, extracted from {\it E. coli}. Error bars show standard deviation.}
\label{fig:Supervised} 
\end{figure}

Figure~\ref{fig:Supervised} shows the prediction accuracies (AUC) for supervised and semi-supervised methods for three different experimental types (knock-out, knock-down and multi-factorial) and the average AUC (all) data. For direct comparison, we included two unsupervised methods (Z-SCORE, SPEARMAN) in our evaluation of supervised methods. Supervised and semi-supervised methods are labeled ``SVM'' followed by the percentage of labeled data (10\%, 30\%, 50\%, 70\%, 90\%, 100\%). The suffix ``$+$'' indicates that only positive data were used and ``$\pm$'' indicates that postive and negative data were used.
For instance, ``SVM-70$\pm$'' describes an SVM trained on 70\% of labeled data (positive and negative). All evaluations are five-fold cross-validated and the complexity parameter $C$ of the SVM was optimized via grid search ($0.1 \dots 100$) for each training fold.

The results show good prediction accuracies for supervised methods on all experimental types, with a slight advantage for knock-out data. As expected, performance increases with the percentage of data labeled but there is little difference between labeling only positive data, or both positive and negative data. Apparently, supervised methods can be trained effectively even when only a portion of network interactions (positives) is known.

Even with as little as 10\% of known interactions, semi-supervised methods still outperform unsupervised methods for multi-factorial data. The Z-SCORE method is still the top-performing method on knock-out data, but supervised methods are not far behind and considerably outperform Spearman's correlation. For knock-down data the Z-SCORE method loses its top rank, and semi-supervised methods perform better when at least 70\% of the data are labeled.

To summarize, apart from the Z-SCORE method on knock-out data, supervised and semi-supervised approaches considerably outperform unsupervised methods and achieve good prediction accuracies in general for networks of this size.

\section{Discussion}

\subsection{Simulated data} 

While simulators such as GenNetWeaver generate expression data that are in good agreement with biological measurements \citep{Marbach2010} they remain incomplete models, e.g. post-transcriptional regulation and chromatin states are missing, and an evaluation of inference methods on real data would clearly be preferable. However, currently known network structures, even for well-characterized organisms, are fragmentary and only partially correct representations of the interactions between genes~\citep{Stolovitzky2007}. Consequently, there is an unknown but probably large discrepancy between the expression data measured and the observed part of the actual network that generates them, rendering assessment of inference methods on observed gene regulatory networks and their expression values very difficult. We therefore have limited our evaluation to {\it in silico} benchmarks, but methods that fail for simulated data are unlikely to succeed in the inference of real biological networks \citep{Bansal2007}.

\subsection{Linear SVMs} 

Another limitation of our study is the choice of linear SVMs for the evaluation of supervised and semi-supervised methods. We prefer linear SVMs over more-powerful non-linear methods for two reasons.
Firstly, linear SVMs are considerably faster to train and have fewer parameters to optimize than non-linear SVMs -- a significant advantage in a comprehensive study.
Secondly, identifying a complex system with many variables (interaction weights) from a small number of samples calls for a simple predictor. We also tried to evaluate transductive SVMs~\citep{Joachims2009} but found them very time-consuming to train, and they achieved accuracies considerably lower than the semi-supervised SVMs (data not shown). We therefore did not perform a full evaluation and do not report results for transductive SVMs.

\subsection{Feature vectors} 

We construct feature vectors by computing the outer product of the expression profiles of two genes. \citet{Cerulo2010} constructed feature vectors by concatenating the two expression profiles. The outer product results in larger feature vectors ($N^2 \ \text{vs.} \ 2 N $) but is independent of the order of the gene pair. The training set is therefore half the size compared with the concatenation approach ($n(n-1)$) and we achieved higher prediction accuracies with the linear SVM. \citet{Cerulo2010}, however, used non-linear SVMs (RBF) that might achieve the same or better accuracies on concatenated feature vectors but are more time-consuming to train and require two parameter ($C$, $\gamma$) to be optimized. It therefore remains an open question, which method is preferable.

SIRENE by \citet{Mordelet2008} takes a different approach, with SVMs trained on feature vectors derived from single profiles. However, it requires knowledge about the transcription factors amongst the genes, and cannot predict interactions between target genes. Since each transcription factor is assigned a separate SVM, feature vectors are of length $N$ and the training set has only $n$ samples, the individual SVMs can be trained very efficiently, but training time is multiplied by the number of transcription factors.

\subsection{Unbalanced data sets} 

Gene regulatory networks tend to be sparse, with the number of positive samples (interactions) typically much smaller than the number of negative samples (non-interactions). Consequently data sets for the training of supervised methods are heavily unbalanced, and this could have a negative impact on the prediction accuracy of the classifier. We therefore tried to weight positive and negative samples inversely to their ratio, but did not observe any improvements in prediction accuracy (data not shown). All evaluations in this paper were therefore performed with equally weighted ($w=1$) samples.

\subsection{Network inference} 

The evaluation results reveal large variations in prediction accuracies across all methods.
Non-linear methods such as MINE do not perform better than linear Pearson's correlation and in general, we find that complex methods are no more accurate than simple methods. The Z-SCORE method and Pearson correlation are the two best-performing unsupervised methods.

\begin{figure}[ht] 
\centerline{\includegraphics[scale=0.80]{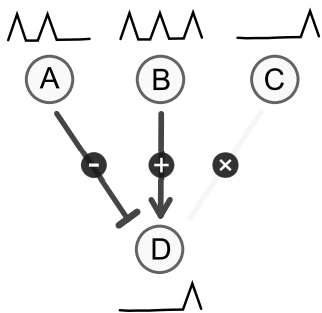}}
\caption{Gene A inhibits gene D and gene B activates gene D. The resulting expression profile of gene D is, however, most similar to that of gene C, which does not regulate gene D.}
\label{fig:FailedInference} 
\end{figure}

A detailed analysis revealed that unsupervised approaches work well for simple network topologies (e.g. star topology) and networks with exclusively activating or inhibiting interactions, but fail for more-complex cases (see Supplemenary Material). Mixed regulatory interactions constitute a fundamental problem for unsupervised network inference as depicted in Figure~\ref{fig:FailedInference}.

Let gene A inhibit gene D but let gene B activate the same gene D. Given the expression profiles of genes A and B as shown in Figure~\ref{fig:FailedInference}, and assuming identical interaction weights but with opposite signs, the  profile for gene D, resulting from a linear combination, is most similar to that of gene C and very different from A or B. Consequently, the most-appropriate but erroneous conclusion is to infer a regulatory relationship between C and D. Without any further information (e.g. knock-outs, existing interactions) any method that infers interactions from the similarity of expression profiles alone is prone to fail in this common case. \citet{Schaffter2011} identify other common network motifs and the methods that tend to infer them incorrectly, and \citet{Krishnan2007} show that networks of a certain complexity cannot be reverse-engineered from expression data alone.

\section{Conclusion}

Perhaps the most-important observation from this evaluation is the large variance in prediction accuracies across all methods. In agreement with \citet{Haynes2009} we find that a large number of replicates on networks of varying size is required for reliable estimates of the prediction accuracy of a method. Evaluations on single data sets -- especially on real data -- are unsuitable to establish differences in the prediction accuracy of inference methods.

On average, unsupervised methods achieve very low prediction accuracies, with the notable exception of the Z-SCORE method, and are considerably outperformed by supervised and semi-supervised methods. Simple correlation methods such as Pearson correlation are as accurate as much more-complex methods, yet much faster and parameterless. Unsupervised methods are appropriate for the inference only of simple networks that are entirely composed of inhibitory or activating interactions but not both.

The Z-SCORE method achieved the best prediction accuracy of all methods on knock-out data, but has obvious limitations. For instance, the method fails when a gene is regulated by an or-junction of two other genes. However, the method could easily be generalized to multi-knock-out experiments.

On multi-factorial data the supervised and semi-supervised methods achieved the highest accuracies; even with as few as 10\% of known interactions, the semi-supervised methods still outperformed all unsupervised approaches. There was little difference in prediction accuracy for semi-supervised methods trained on positively labeled data only, compared to training on positive and negative samples. Apparently semi-supervised methods can effectively be trained on partial interaction data and non-interaction data are not essential. 

These results have important implications for the application of network inference methods in systems biology. Even the best methods are accurate only for small networks of relatively simple topology, which means that large-scale or genome-scale regulatory network inference from expression data alone is currently not feasible. If inference methods are to be applied to data of the scale generated by modern microarray platforms, a feature selection step is usually required to reduce the size of the inference problem; attempts to apply network inference to such large-scale datasets may be premature, and consideration should be given to focusing the biological question to use smaller-scale, higher-quality experimental data. 

Our analysis also indicates that certain kinds of biological data are more amenable for accurate network inference than others. Most microarray datasets are most similar to our multi-factorial simulations, which yielded poorly inferred networks with unsupervised methods. Increasing the number of samples in the experiment (a common strategy to improve inference) does not in fact generate the hoped-for improvements. More useful are knock-out data, which our simulations show contain more-useful information, and support higher-quality inference. Biologists who wish to gain insight into regulatory architecture should consider these limitations when designing experiments.

To summarize, small networks (as evaluated here) can be inferred with high accuracy (AUC $\approx$ 0.9) even with small numbers of samples using supervised techniques or the Z-SCORE method. However, even with the best-performing methods large variations in prediction accuracy remain, and predictions may be limited to undirected networks without self-interactions.

\section*{Funding}

We acknowledge funding from the Australian Research Council DP110103384 and CE0348221.

\bibliography{GRNI}{}
\bibliographystyle{plainnat}

\end{document}


\title{
  {\bfseries Supplementary material } \\ 
  \vspace{50pt}
  Supervised, semi-supervised and unsupervised inference of gene regulatory networks 
}   
       
\author{Stefan R. Maetschke, Piyush B. Madhamshettiwar, Melissa J. Davis and Mark A. Ragan}        

\maketitle

\pagebreak
\section{Unsupervised}

This sections contains additional data of unsupervised methods for different performance metrics
and experimental data types.

\subsection{Methods}

The following three figures show the prediction performance of unsupervised methods for three different performance measures such as the Area Under the ROC curve (AUC), Matthew's Correlation Coefficient (MCC) and the F1-score. The threshold for the MCC and F1 score metrics were optimized. The AUC does not have a threshold that requires optimization. 

All methods were evaluated on multi-factorial, knock-out, knock-down
and averaged (all) data generated by GenNetWeaver. Each evaluation was repeated 10 times over networks with 10,...,110 nodes, extracted from {\it E. coli} and {\it S. cerevisiae} networks.

\begin{figure}[H] 
\centerline{\includegraphics[angle=0,scale=0.7]{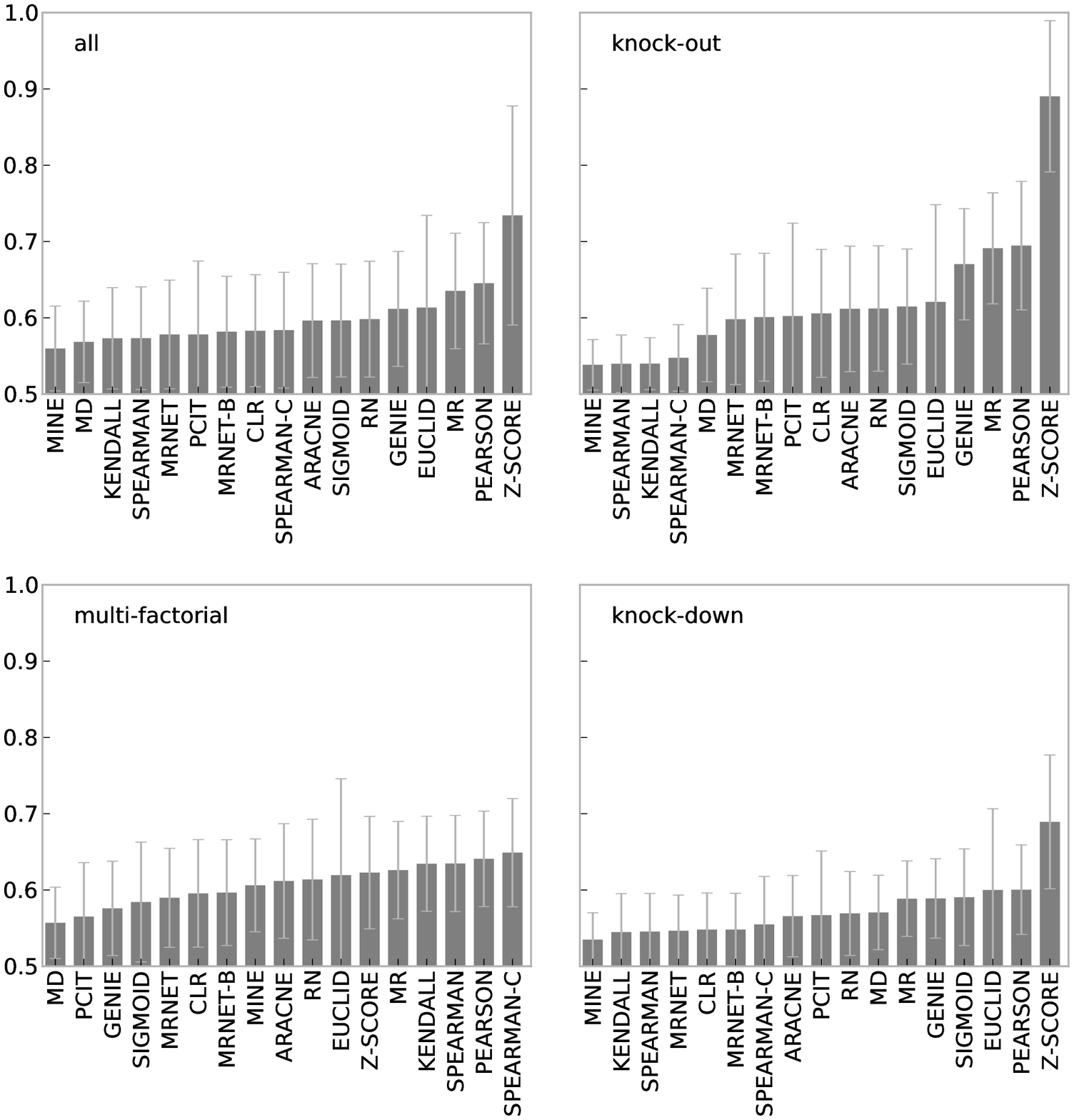}}
\caption{Prediction accuracy (AUC) of unsupervised methods for different experimental data types. Error bars show standard deviation.}
\label{fig:Unsupervised_AUC} 
\end{figure}

While there are slight differences in the ranking of the methods depending on the chosen performance metric no dramatic shifts can be observed. Z-SCORE and PEARSON remain the best performing methods in all cases and the Z-SCORE method dominates all other methods for knock-out data.

\begin{figure}[H] 
\centerline{\includegraphics[angle=0,scale=0.7]{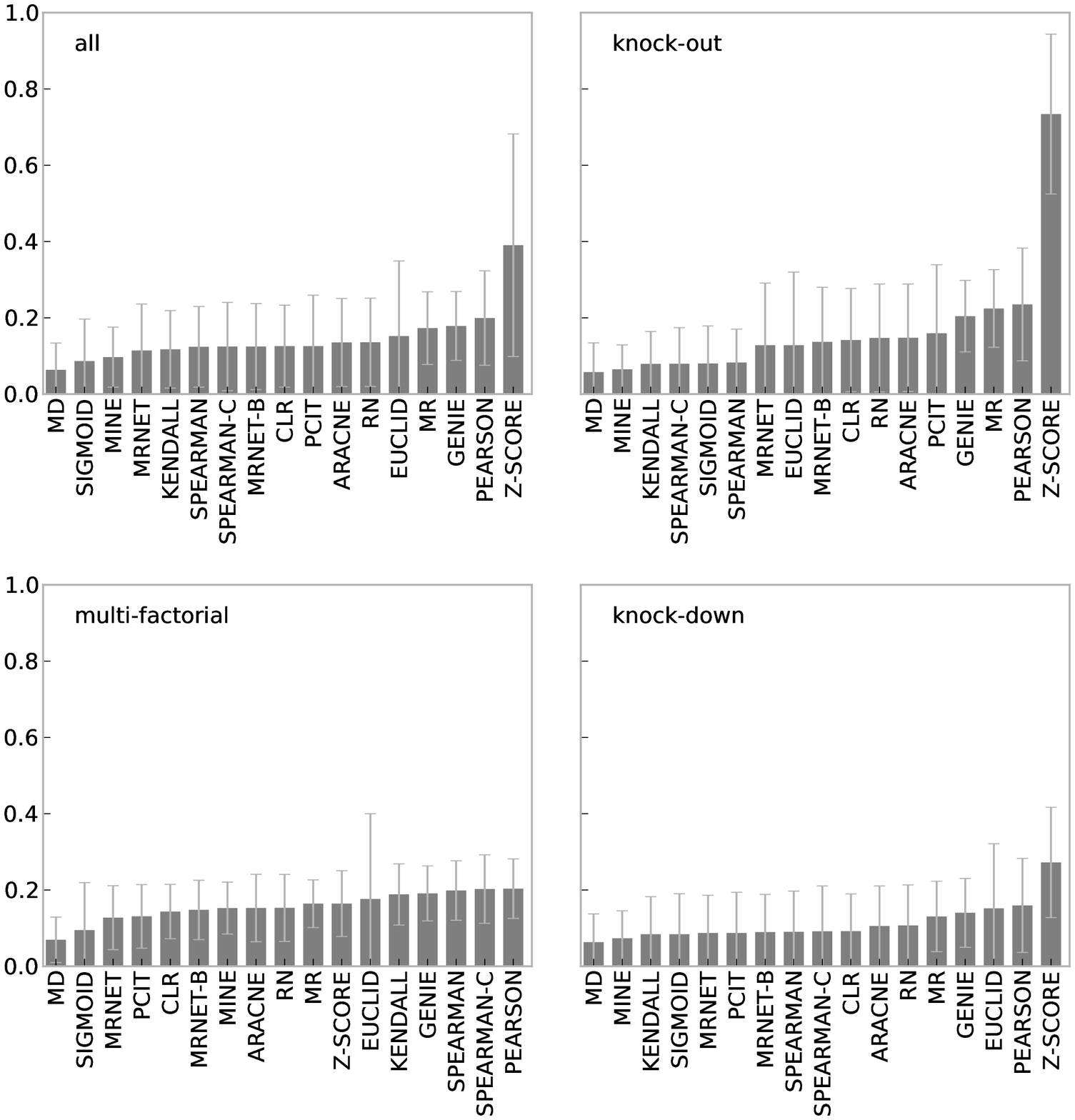}}
\caption{Prediction accuracy (MCC) of unsupervised methods for different experimental data types. Error bars show standard deviation.}
\label{fig:Unsupervised_MCC} 
\end{figure}

\begin{figure}[H] 
\centerline{\includegraphics[angle=0,scale=0.7]{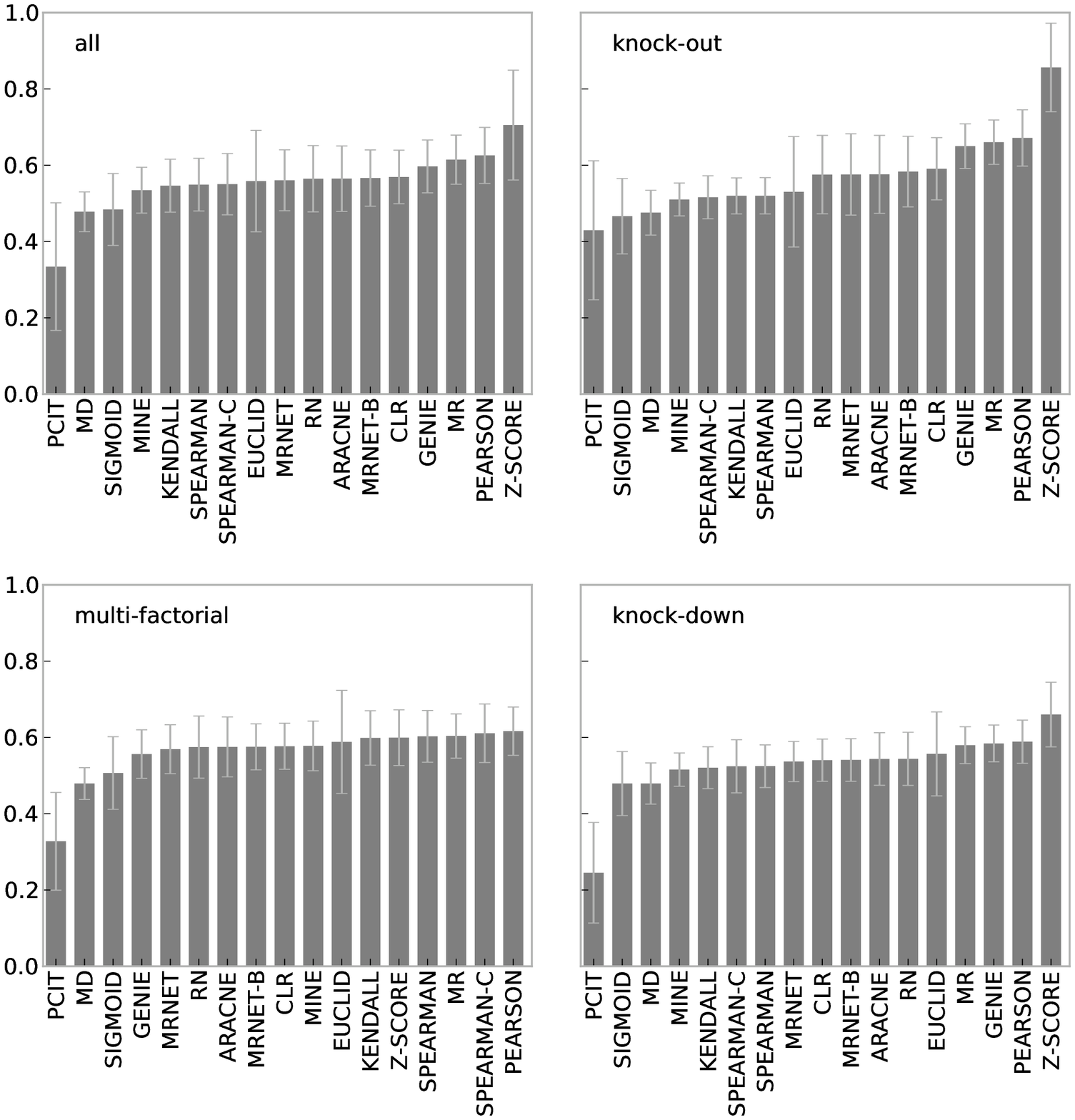}}
\caption{Prediction accuracy (F1-score) of unsupervised methods for different experimental data types. Error bars show standard deviation.}
\label{fig:Unsupervised_FSCORE} 
\end{figure}

\pagebreak
\subsection{Network size}

This section shows the prediction performance (AUC) of the unsupervised methods for networks with different node numbers and for the three experimental types (multi-factorial, knock-down, knock-out). All expression data were simulated with GenNetWeaver and sub-networks were extracted from {\it E. coli}.

\begin{figure}[H] 
\centerline{\includegraphics[scale=0.7]{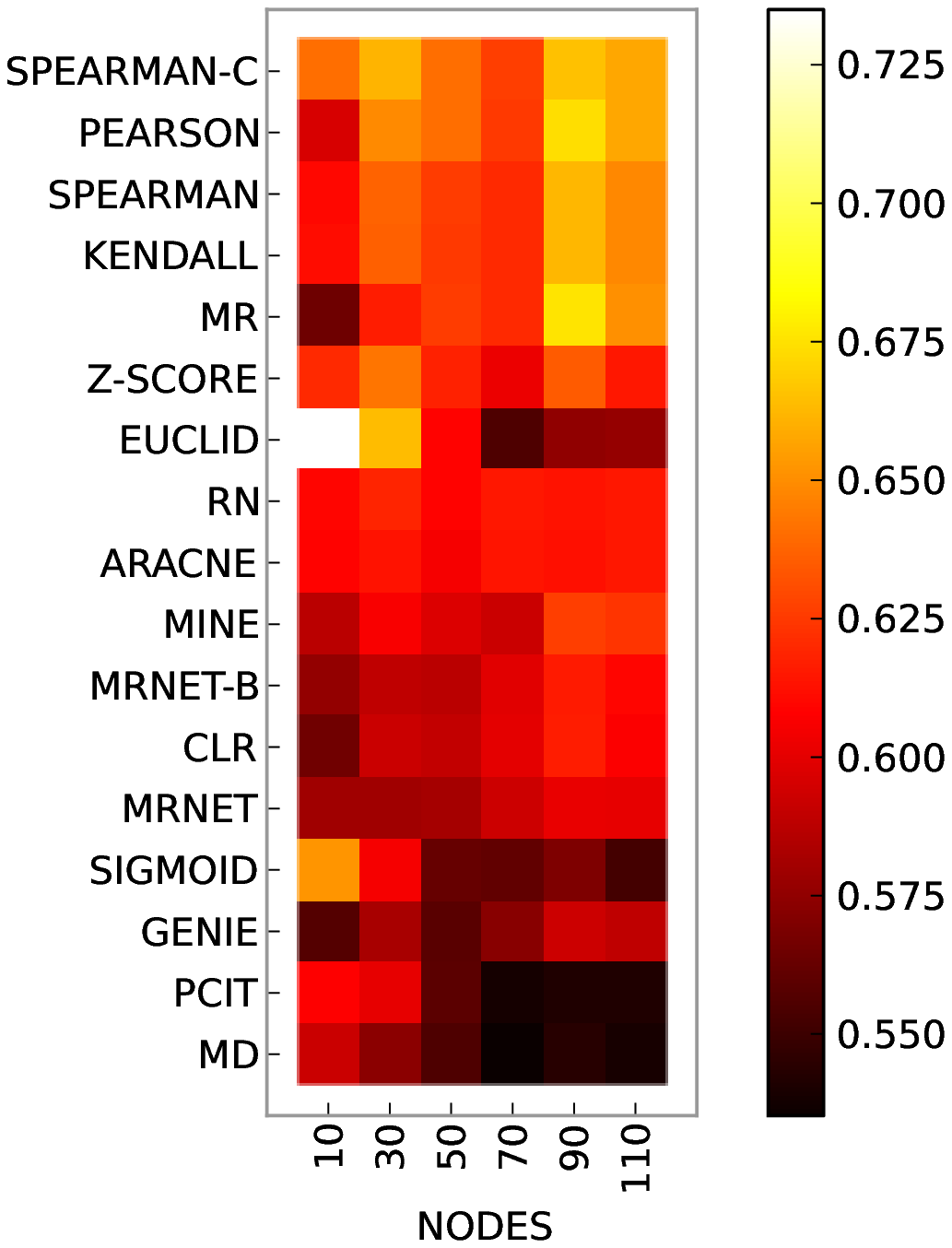}}
\caption{Prediction accuracy (AUC) of unsupervised methods on multi-factorial data for different network sizes. }
\label{fig:Unsupervised_multi} 
\end{figure}

Figure~\ref{fig:Unsupervised_multi} reveals that the best performing unsupervised method on multi-factorial data is the EUCLID method but only on very small networks with 10 to 30 nodes. Correlation based methods such as PEARSON, SPEARMAN-C, SPEARMAN, KENDALL and some other methods show better performance on larger networks (90 and 110 nodes) than on smaller networks.

\pagebreak

\begin{figure}[H] 
\centerline{\includegraphics[scale=0.7]{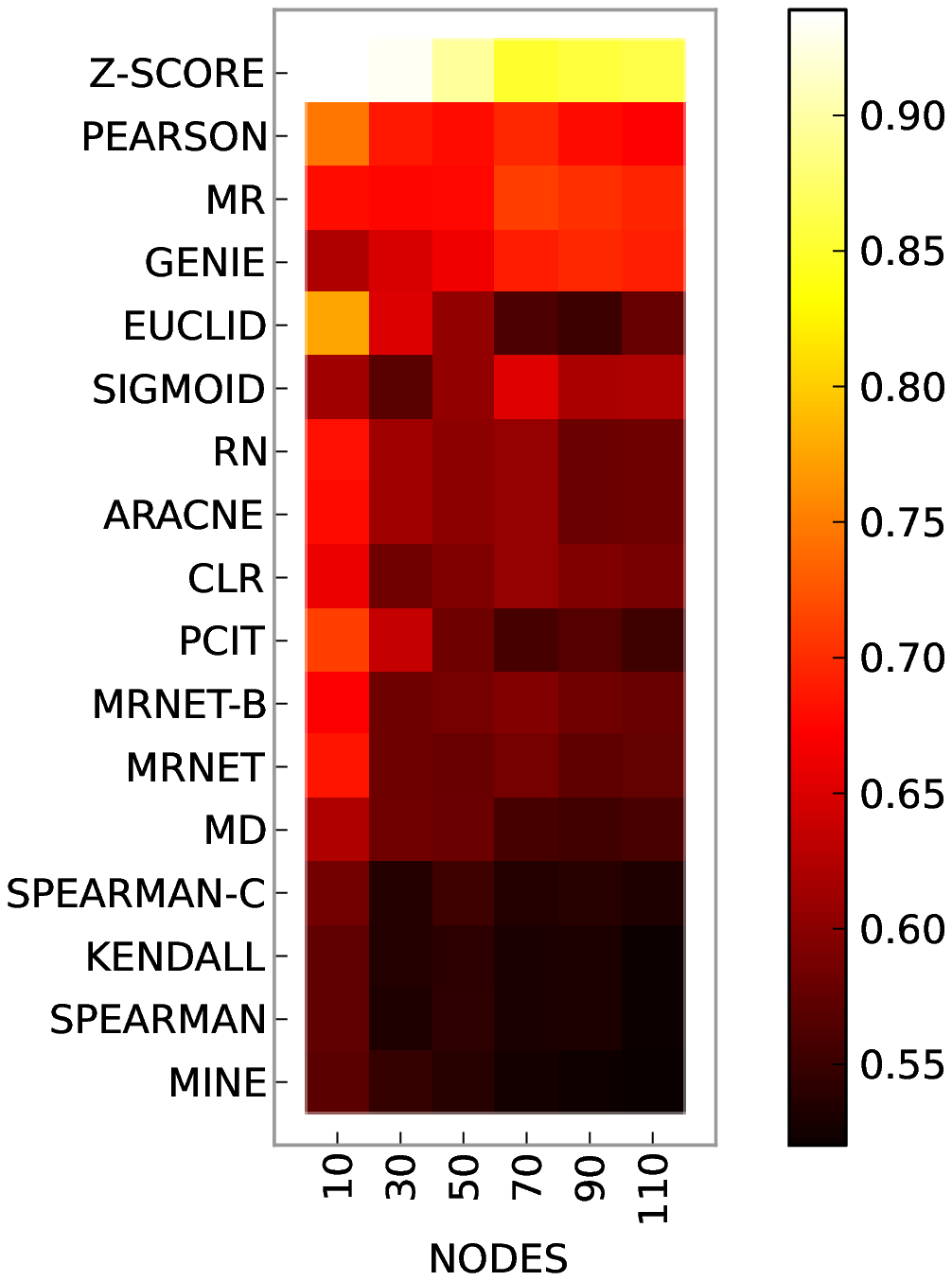}}
\caption{Prediction accuracy (AUC) of unsupervised methods on knock-out data for different network sizes. }
\label{fig:Unsupervised_knockout} 
\end{figure}

On knock-out data the most accurate method is the Z-SCORE method. While the prediction accuracy of the Z-SCORE method decreases with network size it still clearly outperforms all other methods for networks of all sizes (see Figure~\ref{fig:Unsupervised_knockout}). There is a general trend for most methods to perform better on the small 10-node network. Apart from PEARSON, all correlation based methods (SPEARMAN-C, SPEARMAN, KENDALL, MINE) achieve very low AUCs on knock-out data.

\pagebreak

\begin{figure}[H] 
\centerline{\includegraphics[scale=0.7]{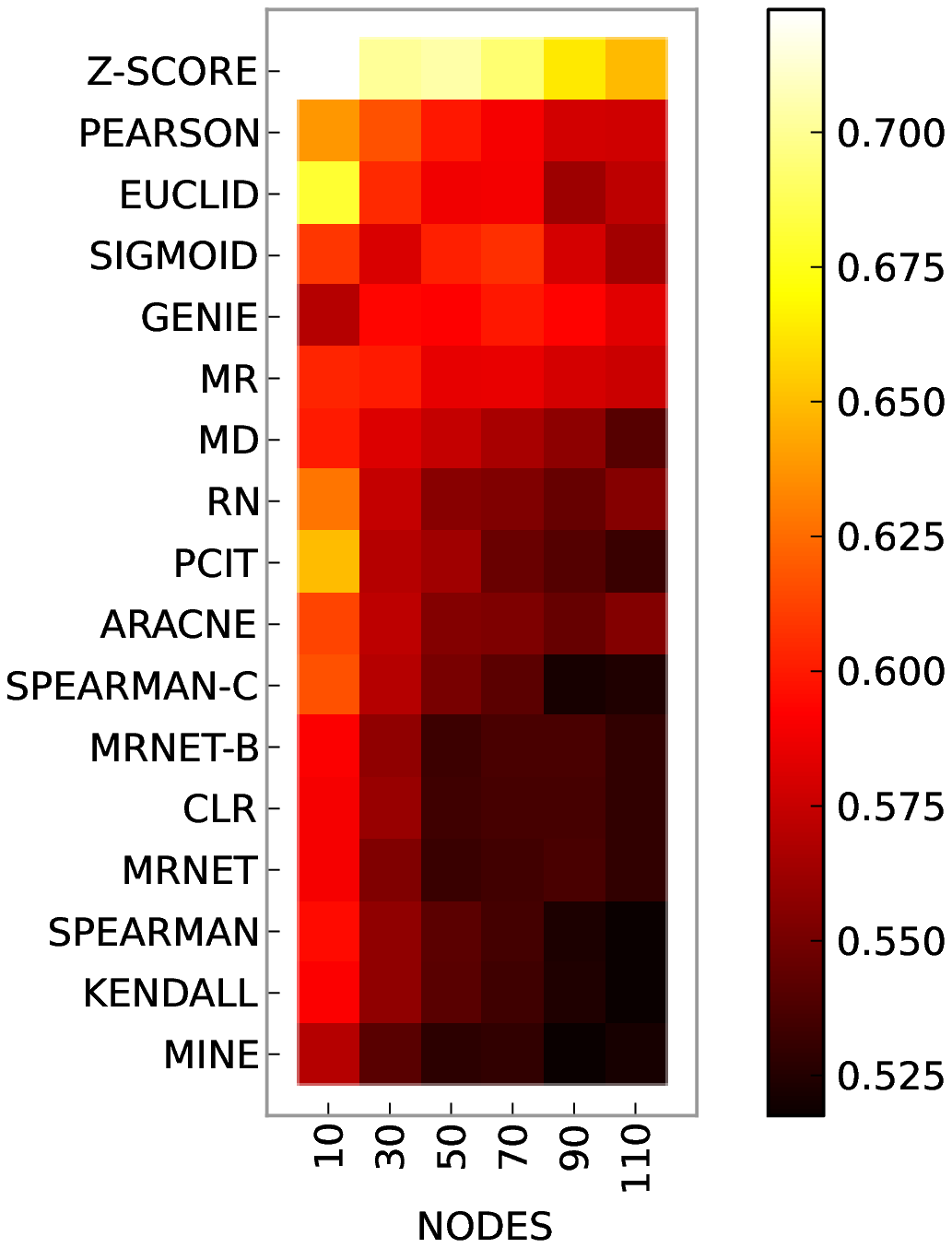}}
\caption{Prediction accuracy (AUC) of unsupervised methods on knock-down data for different network sizes. }
\label{fig:Unsupervised_knockdown} 
\end{figure}

The results on the knock-down data shown in Figure~\ref{fig:Unsupervised_knockdown} are similar to the results on the knock-out data (see Figure~\ref{fig:Unsupervised_knockout}). The Z-SCORE method remains the best performing method. The large majority of methods perform best on the small 10-node network -- especially the EUCLID method, which was the best performer on networks of this size for multi-factorial data.

\pagebreak
\subsection{Network predictions}

All evaluation showed large variations in the prediction accuracy of the methods. Even for very small networks with only 10 nodes the prediction accuracy can vary from perfect to completely wrong. To better understand the reasons causing the large variances we visualized the networks (out of 100) that were predicted with the highest and lowest accuracy, using Spearman's correlation as a network inference method and the AUC as performance metric. 
Sub-networks with 10 nodes were extracted from the {\it E. coli} network and expression data were simulated with GenNetWeaver.

\begin{figure}[H] 
\centerline{\includegraphics[scale=0.7]{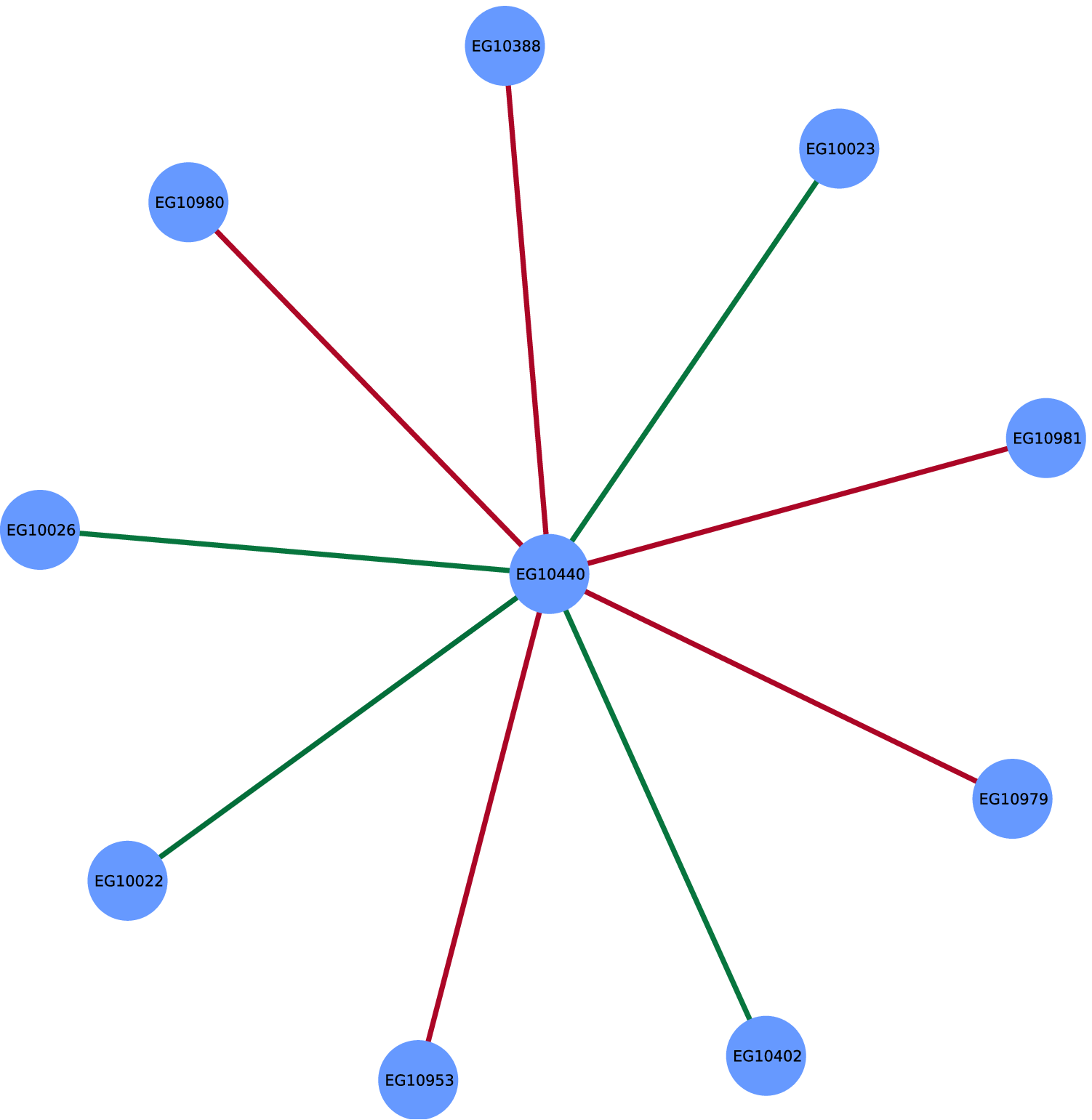}}
\caption{True network where Spearman's correlation failed to recover the topology (AUC = 0.508). Green means activating, and red means inhibiting interactions}
\label{fig:SpearmanNetworkWorst} 
\end{figure}

\begin{figure}[H] 
\centerline{\includegraphics[scale=0.7]{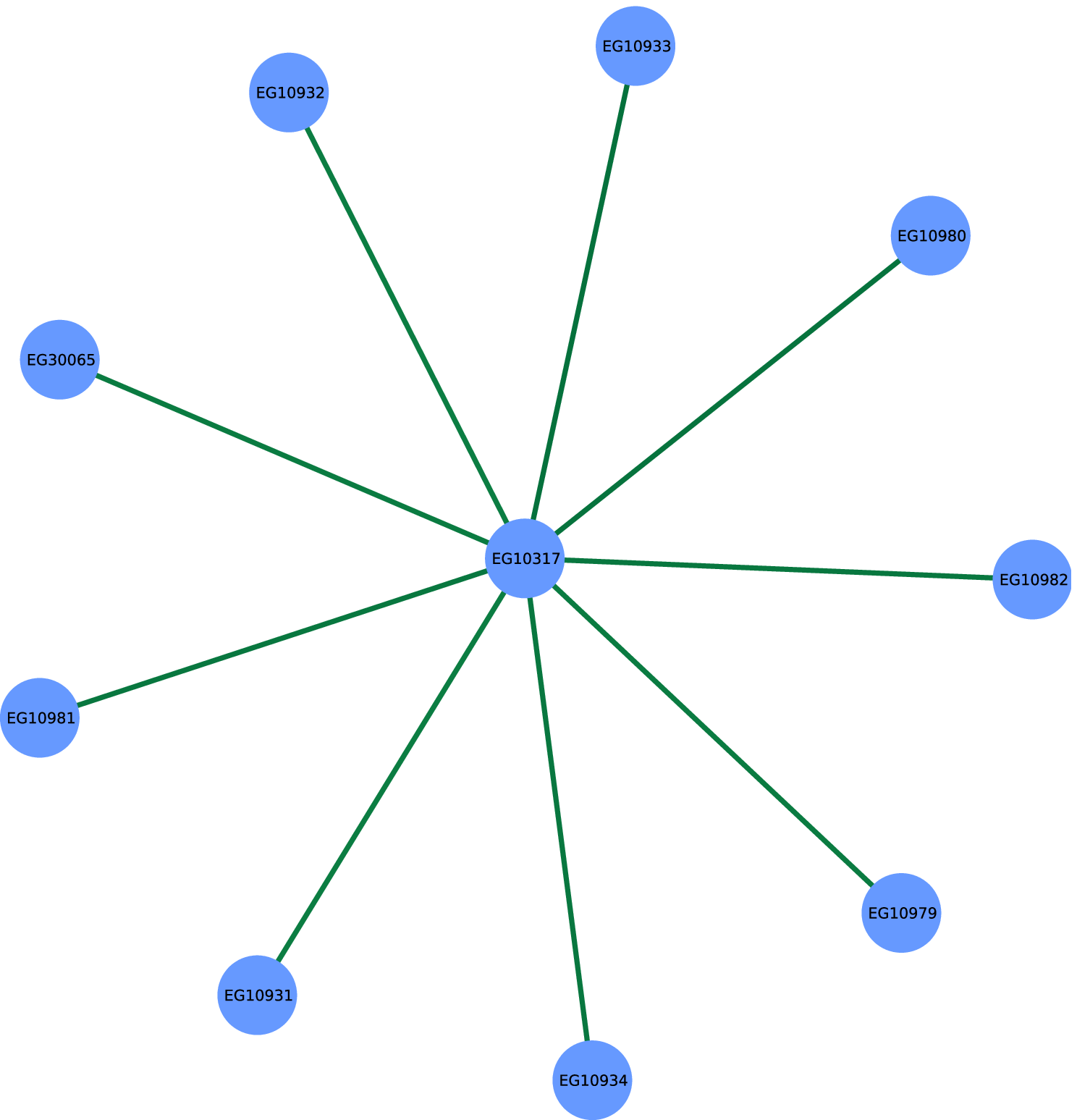}}
\caption{True network where Spearman's correlation recovered the topology accurately (AUC = 0.971). The network has only activating (green) interactions.}
\label{fig:SpearmanNetworkBest} 
\end{figure}

Figure~\ref{fig:SpearmanNetworkWorst} shows a true network where Spearman's correlation failed to recover the topology (AUC = 0.508). Note that some interactions are activating (green) and some interactions are inhibiting (red), which results in a more complex dynamic of the network than a network with exclusively activating or inhibiting interactions.
In contrast, Figure~\ref{fig:SpearmanNetworkBest} shows the true network where Spearman's correlation inferred the network topology close to perfect (AUC = 0.971).

In general, networks with exclusively activating or inhibiting interactions and simple topologies (e.g. star topology) can be inferred accurately with unsupervised methods, even on multi-factorial data. However, networks with complex topologies or a mix of activating and inhibiting interactions typically cannot be recovered reliably from multi-factorial data.

\pagebreak
\section{Supervised}

This section compares supervised, semi-supervised and unsupervised methods, using three different performance metrics such as the Area Under the ROC curve (AUC), Matthew's Correlation Coefficient (MCC) and the F1-score.

All methods were evaluated on multi-factorial, knock-out, knock-down and averaged (all) data generated by GenNetWeaver. 5-fold cross-validation was used and each evaluation was repeated 10 times over networks with 30 nodes, extracted from {\it E. coli}.

The results show little difference in the ranking of the methods for different performance metrics. The Z-SCORE method achieves the highest accuracies on the knock-out data but performs worst on multi-factorial data. SPEARMAN typically shows the lowest prediction accuracy and semi-supervised methods are effectively ranked according to the percentage of labeled data used. No distinction between semi-supervised methods trained on positives and negatives and methods trained on positives-only can be observed.

\begin{figure}[H] 
\centerline{\includegraphics[angle=0,scale=0.7]{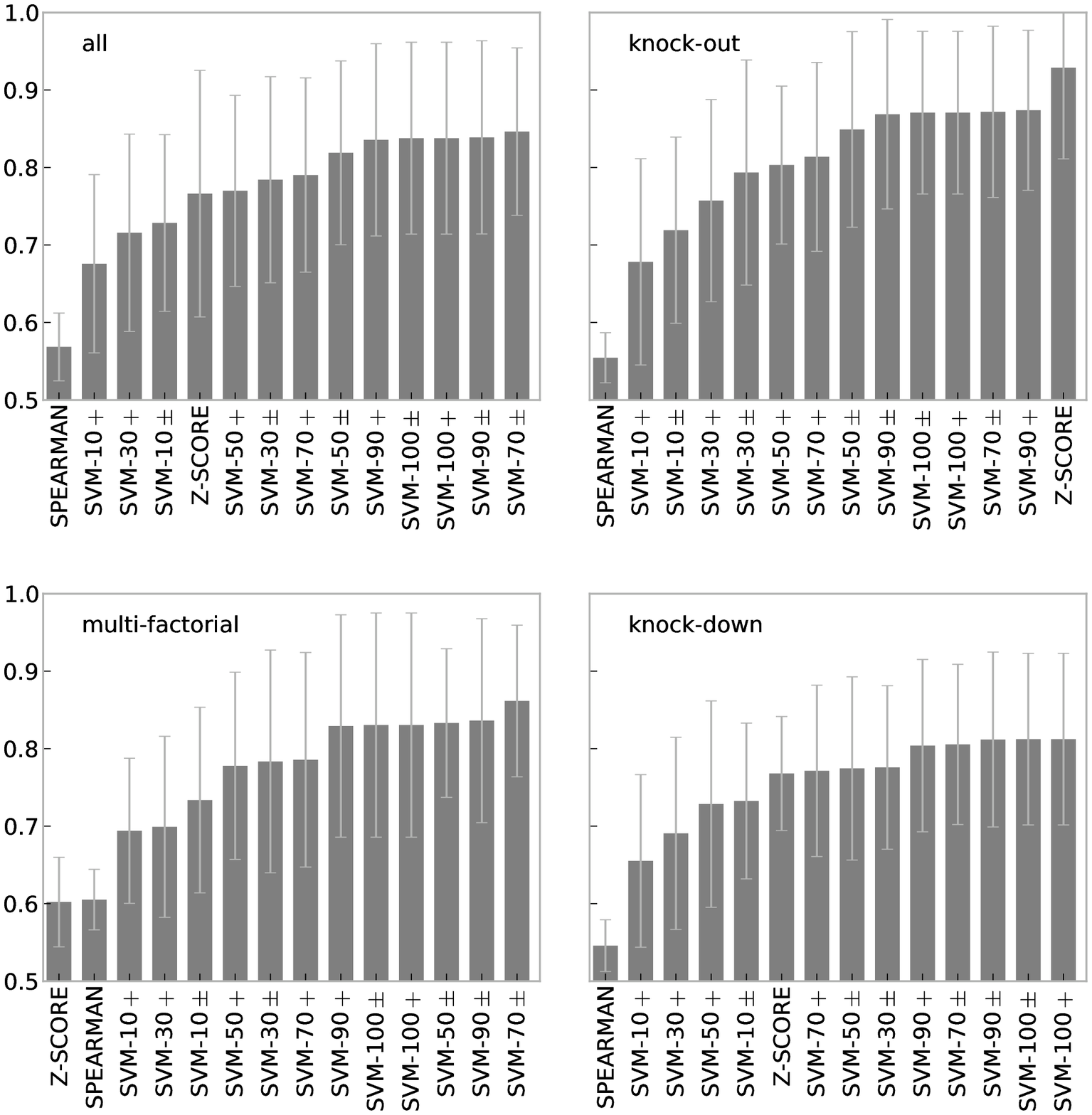}}
\caption{Prediction accuracy (AUC) of supervised methods on multi-factorial, knock-out, knock-down
and averaged (all) data generated by GenNetWeaver.  Error bars show standard deviation.}
\label{fig:Supervised_AUC} 
\end{figure}

\begin{figure}[H] 
\centerline{\includegraphics[angle=0,scale=0.7]{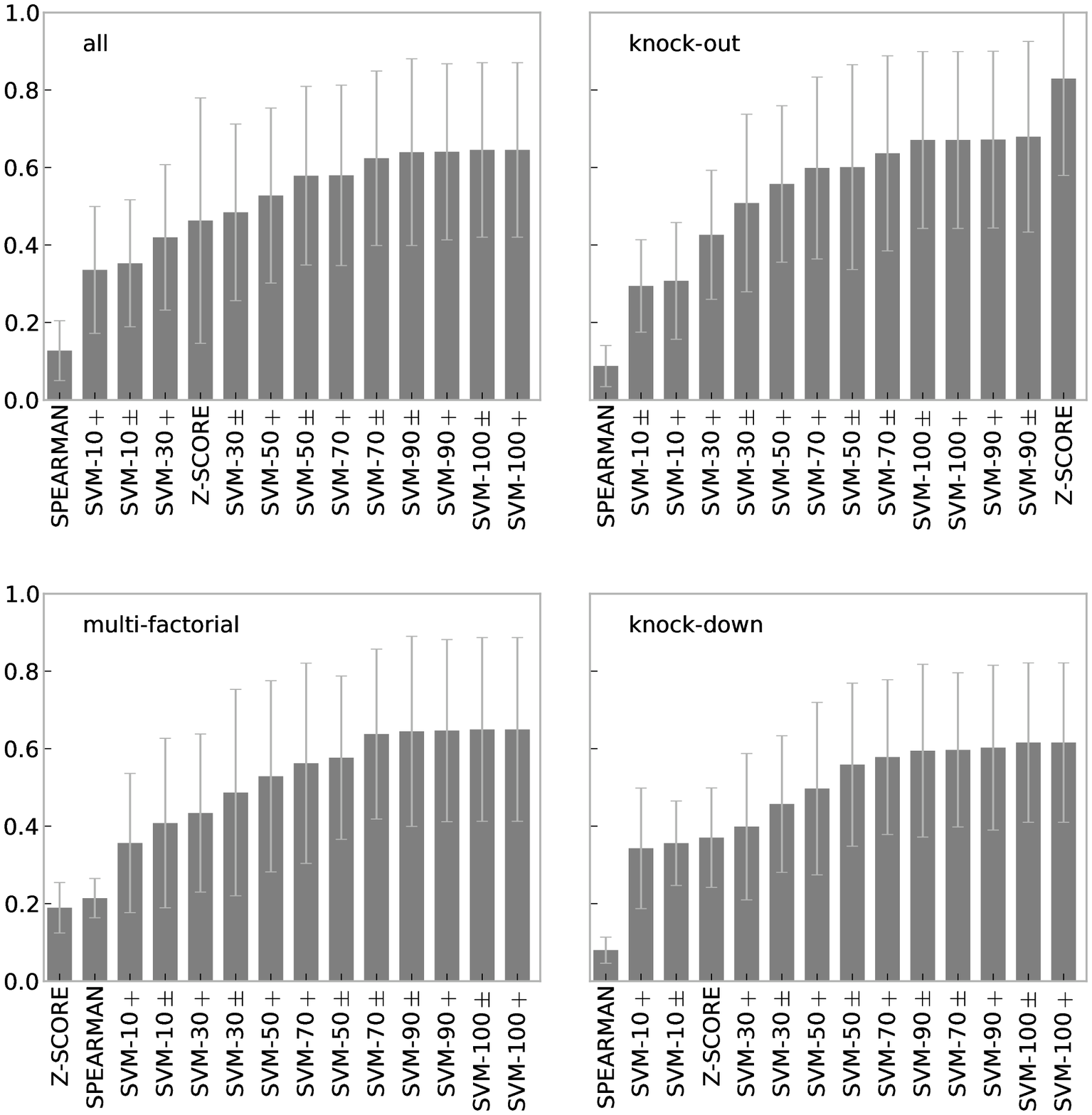}}
\caption{Prediction accuracy (MCC) of supervised methods on multi-factorial, knock-out, knock-down
and averaged (all) data generated by GenNetWeaver. Error bars show standard deviation.}
\label{fig:Supervised_MCC} 
\end{figure}

\begin{figure}[H] 
\centerline{\includegraphics[angle=0,scale=0.7]{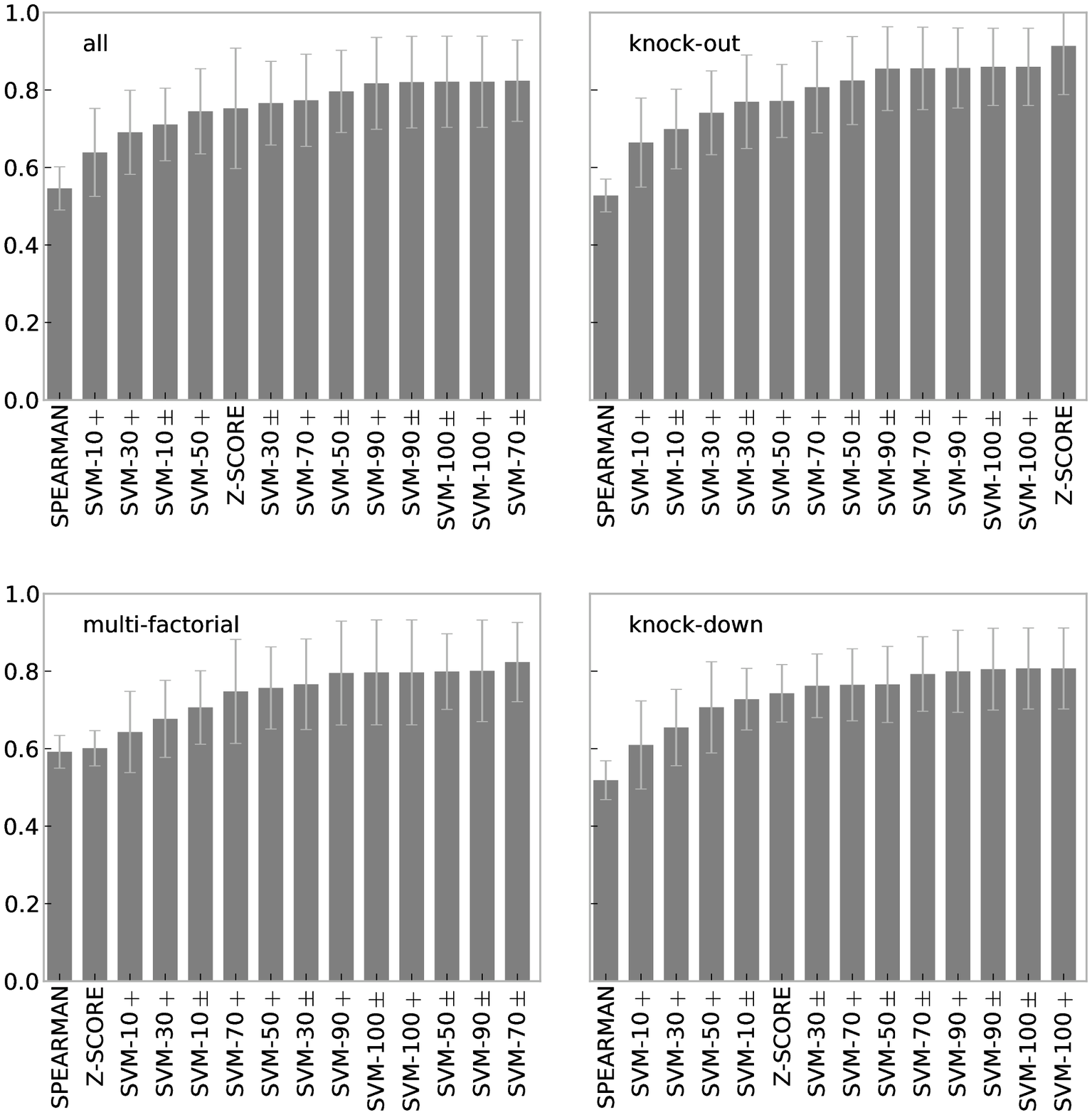}}
\caption{Prediction accuracy (F1-score) of supervised methods on multi-factorial, knock-out, knock-down
and averaged (all) data generated by GenNetWeaver.  Error bars show standard deviation.}
\label{fig:Supervised_FSCORE} 
\end{figure}